# Homology, equilibrium, and conservation laws I:
# Discrete systems of points

D. H. Delphenich
Spring Valley, OH 45370

**Abstract:** The methods of abstract simplicial homology and cohomology are reviewed and applied to the topology of electrical networks. Kirchhoff's laws of electrical circuits are shown to be manifestly homological in their origins. Since they are based in conservation laws, the geometric realization of abstract simplicial complexes is then reviewed and applied to the case of mechanical networks. The equilibrium condition for statics, the conservation laws for closed systems, and the balance principles for open systems are then shown to admit homological formulations.

# Contents

**1. Introduction.** – Although nowadays the branch of mathematics that goes by the name of "homology theory" is often treated as an advanced topic of topology or perhaps a branch of algebra, in the form of homological algebra, nonetheless, in the eyes of history, the roots of homology were established long before Felix Haussdorff defined the point-set theoretic axioms of a topology on a set that are used nowadays in his 1914 classic *Grundzüge der Mengenlehre* (*Basic Set Theory*) [**1**].

One of the earliest manifestations of homological methods was due to the Swiss mathematician Leonhard Euler, whole solved the famous "Königsberg Bridge problem" in 1736 by using what would later become the "Euler-Poincaré characteristic" of the space in question, which he defined in a 1752 paper. That problem concerned the possibility of defining what would now be called an "Eulerian cycle" or "Hamiltonian cycle" for the graph that is defined by an island in the Preger river that goes through Königsberg (which was then in Prussia, but is now known as Kaliningrad, which is in Russia) and the seven bridges that connected it with the banks of that river. Such a path would have to cross each bridge exactly once and return to its starting point. What Euler showed was that although an Eulerian path could exist, an Eulerian cycle could not. The people who work in graph theory also regard this as the start of their own discipline, but it is important for what follows to see that a graph can also be regarded as a one-dimensional "abstract simplicial complex," which is what will be treated in what follows.

August Ferdinand Möbius, who was just as distinguished in the world of structural analysis and graphical statics as he was in geometry, devised what he called the "barycentric calculus" in 1844 [**2**]. That calculus actually contained many of the elements of what later became known as "combinatorial topology."

Another big advance in homology or its specialization to graph theory was made a year later by Gustav Kirchhoff (who was born in Königsberg) in 1845 [**3**]. While still a student at the Albertus University of Königsberg, Kirchhoff attended a seminar on mathematical physics that was given by Carl Gustav Jacob Jacobi, Franz Ernst Neumann, and Friedrich Julius Richelot, and as an exercise, he proved what are now referred to as "Kirchhoff's laws of electrical circuits"; that later became the basis for his doctoral dissertation. Those laws say that the sum of the currents at any node of an electrical network will vanish, as will the sum of the voltage drops around any loop.

Gradually, the methods that were being used for the aforementioned problems gave way to the branch of geometry that came to be called *analysis situs*, which is Latin for "the analysis of position," just as "topology" comes from the Greek roots *topos* and *logos*, which also amount to the analysis of position. However, in retrospect, the only role that the "position" of a region in space played in all of that was that one was looking for properties of the region that were *independent* of position. One of the great works on analysis situs was a series of papers by Henri Poincaré that started in 1895 [**4**], in which he expanded the definition of the Euler characteristic to higher dimensions, as well as posing his celebrated conjecture about three-dimensional differentiable manifolds. Interestingly, his definition and treatment of differentiable manifolds actually predated Haussdorff's set-theoretic definition of a topology by two decades. One of the later classics in analysis situs was first published by Oswald Veblen in 1921 [**5**].

Meanwhile, the term "analysis situs" was gradually giving way to *combinatorial topology*, which started with the set-theoretic definition of a topology and attempted to use the basis building blocks of analysis situs – namely, *simplexes* of various dimensions

– as a way of representing a topological space by a *simplicial complex*. One of the earliest classics of that approach was called *Topologie*, which was published by Paul Alexandroff and Heinz Hopf in 1936 ([**6**], see also [**7**] and [**8**]).

The spirit of the present monograph goes back to some posthumously-published lecture notes of Solomon Lefschetz on applications of algebraic topology [**9**], the first part of which was devoted to the application of homology to electrical networks. It is interesting that Lefschetz was originally educated an electrical engineer in Russia, but lost both forearms to a transformer explosion in a factory, and then redirected his interests towards topology. In fact, he cited the book by Veblen on analysis situs as a major source of inspiration to focus on topology. It is also interesting that other topologists with links to electrical networks include the late Raoul Bott, whose doctoral dissertation in applied mathematics was concerned with a mathematical problem in the theory of electrical networks, and Victor Guillemin, whose brother Ernst was distinguished by his contributions to the theory of networks.

Although homology theory was rooted in applications, as is often the case in mathematics, the tendency to generalize and abstract the mathematical concepts that were originally important in the applications created an increasingly esoteric branch of pure mathematics in the form of algebraic topology and homological algebra, which abstracts the algebraic techniques from the topological ones. As a result, many researchers outside of that abstract realm are quite reluctant to consider it as a set of tools for mathematical model building and problem solving for problems that are rooted in more practical considerations, such as science, engineering, and economics. That has had the effect of largely eliminating the possibility of any applications from an otherwise purely-mathematical discipline. In physics, one finds more applications of homotopy theory to the study of topological defects in ordered media, even though the place of homotopy in that study is that of a local coefficient group for the obstruction cocycles that relate to defining the relevant fields globally (see the author's paper on that topic [**10**]). Certainly gauge field theories are a major consumer of the methods of characteristic classes, which relate to the cohomology of fiber bundles. However, the main applications for homology theory outside of the aforementioned domains are largely based in its specialization to the topology of networks or graphs.

The present work is basically an expansion upon the fact that it is no coincidence that Kirchhoff's laws of electrical circuits are based upon conservation laws. In particular, the law of currents comes from the conservation of charge, and the law of voltages comes from the conservation of energy. Furthermore, the former law can be expressed by saying that current defines a "1-cycle" in the abstract simplicial complex that the network defines, while the latter one says that voltage drops define a "1-cocyle."

One finds that analogous results are to be found in the study of mechanical networks, such as structures, trusses, frameworks, and the like. In that context, one can also give a homological basis for the concept of equilibrium in statics, as well as the balance principles of dynamics. Indeed, the concept of work is also deeply rooted in homological notions. The overarching general idea that should emerge from what follows is that the most fundamental first principles of physics, namely, the conservation laws for closed systems and balance principles for open ones, have their roots in homology and cohomology.

The first part of this series of monographs is devoted to finite systems of interacting points, which can then be modeled by graphs or one-dimensional simplicial complexes. The second part of the series will then go on to dimensions that are higher than one in order to discuss continuous systems of points. The homology that is introduced in the process will be actually be quite minimal, so only a passing familiarity with point-set topology and abstract algebra will be expected. Typically, the results of the definitions will take the form of identifying various physical notions as being boundaries or coboundaries, cycles or cocycles. The introduction of homology into the applications will also be done in two stages: First, abstract simplicial homology will be introduced in order to discuss Kirchhoff's laws of circuits, and then the geometrical realizations of abstract simplicial complexes in affine spaces will be introduced in order to discuss mechanical networks.

## 2. Abstract simplicial homology.

**2. Abstract simplicial homology.** The type of homology that is most naturally adapted to the concerns of most physical networks is "abstract simplicial homology [**11**]." Hence, we shall attempt to summarize the essential notions of that subject to the extent that they are relevant to the physical models under consideration.

*a. Abstract simplicial complexes.* – The basic building blocks for an *abstract simplicial complex* consist of abstract simplexes of successive dimensions. For our purposes, all of the relevant sets shall be finite sets.

We start with a set $S_0 = \{i_1, \ldots, i_{r_0}\}$ that we call the *vertex system* of the complex; we shall also refer to it as the *0-skeleton* of that complex. We shall also often refer to a vertex as a *node*. Our choice of notation for a vertex at this point is based in the idea that mostly it is defined by an integer value for an index.

In the next dimension, the *1-simplexes* of this abstract simplicial complex are then a specified subset $S_1$ of $S_0 \times S_0$; that is, a specified collection of pairs $(i, j)$ of indices in $V$. We shall often regard the pair $(i, j)$ as a "branch" of the "network" that we are defining when the highest order of simplexes that we are using is one. It is important to note that since the elements of $S_1$ are ordered pairs of vertices, the pair $(i, j)$ will be distinct from the pair $(j, i)$, which might not even belong to $S_1$ anymore. We shall then think of $(i, j)$ and $(j, i)$ as the two *orientations* of the same branch.

Next, the *2-simplexes* are a specified subset $S_2$ of triples $(i, j, k)$ of distinct indices. One can then continue in this manner and define *p*-simplexes in each dimension *p*. One then defines the *dimension* of the complex to be the highest dimension of its simplexes.

By recursion, the *p-skeleton* $\mathcal{S}_p$ of the abstract simplicial complex is defined to be the union $S_0 \cup \ldots \cup S_p$. When *n* is the dimension of the complex, $\mathcal{S}_n$ will then be referred to as simply the (abstract) complex.

*b. Abstract k-chains* ([1]). – Let *R* be a commutative ring with unity (which will be denoted by 1). Hence, multiplicative inverses to scalars do not have to exist (except for 1). Note that a ring with unity that is infinitely cyclic must include the ring of integers $\mathbb{Z}$, since it must define an Abelian group that will include all finite sums of 1 with itself $m \equiv 1 + \ldots + 1$, as well as its additive inverse $-1$. If it is finitely cyclic then there will be a maximum $m$ for which $m + 1 = 0$. We shall not be concerned with finitely-cyclic rings here, however.

---

([1])  For more details on the concepts of rings and modules, see, e.g., Birkhoff and MacLane [**12**].

If the elements of $R$ are regarded as coefficients then the simplexes in each dimension can be used as the generators for what one calls *free R-modules*. Such an algebraic structure is essentially a vector space whose scalars come from a ring $R$ that does not have to be a field, such as $\mathbb{Z}$, the ring of integer; A free $R$-module of rank $r$ will then take the form of $R^r = R \oplus \ldots \oplus R$, with $r$ summands. Hence, it is the closest analogue of the vector space $\mathbb{R}^r$ in terms of $R$-modules.

If we now denote a $k$-dimensional simplex by $\sigma_k$, so the set of all $k$-dimensional simplexes will take the form $\{\sigma_k(i), i = 1, \ldots, r_k\}$, then an element of the free $R$-module $C_k(\mathcal{S}_n ; R)$ of rank $r_k$ that has that that set of simplexes for its generators will take the form of a formal sum:

$$c_k = \sum_{i=1}^{r_k} a(i)\,\sigma_k(i)\,. \qquad (2.1)$$

Such a formal sum is referred to as an *abstract k-chain with coefficients in R.*

Although the concept of a formal sum can be made mathematically rigorous, for the sake of computation, it is usually sufficient to treat it as a set of rules for symbol manipulation. Hence, to be consistent, if $m$ is a positive integer then one must have:

$$m\,\sigma_k = \sigma_k + \ldots + \sigma_k \quad (m \text{ summands}), \qquad (-1)\,\sigma_k = -\,\sigma_k\,. \qquad (2.2)$$

As a consequence: (1) $\sigma_k = \sigma_k$.

In order to interpret (0) $\sigma_k$, one must think of multiplication by 0 as eliminating that term from the sum, and the "empty sum" is represented by 0. Since the empty set is a subset of any set, 0 is a $k$-chain for every $k$. Hence, one can define the addition and subtraction of $k$-chains by implicitly including summands of the form (0)$\sigma_k = 0$, if necessary. If $c_k$ is as in (4.14) and $c'_k$ is defined analogously then:

$$c_k + c'_k = \sum_{i=1}^{r_k} m(i)\,\sigma_k(i) + \sum_{i=1}^{r_k} m'(i)\,\sigma_k(i) = \sum_{i=1}^{r_k} [m(i) + m'(i)]\,\sigma_k(i)\,, \qquad (2.3)$$

which will then make:

$$c_k - c'_k = c_k + (-1)\,c'_k = \sum_{i=1}^{r_k} [m(i) - m'(i)]\,\sigma_k(i)\,, \qquad c_k - c_k = (1-1)\,c_k = 0. \qquad (2.4)$$

Hence, "scalar" multiplication by an element of $R$ must distribute over the formal addition of simplexes.

The direct sum of all such $R$-modules up to the dimension of $\mathcal{S}_n$ will be an $R$-module that will be denoted by:

$$C_* \left( \mathcal{S}_n \, ; R \right) = C_0 \left( \mathcal{S}_n \, ; R \right) \oplus \, \ldots \, \oplus \, C_n \left( \mathcal{S}_n \, ; R \right) .$$

Its elements will be finite formal sums of chains of varying dimensions.

*c. The abstract boundary operator.* – For the sake of brevity, it is convenient to denote $k$-simplexes as "words" formed from the "letters" of the "alphabet" $S_0$ . For instance, the 1-simplex $(i, j)$ is denoted by $ij$, the 2-simplex $(i, j, k)$ by $ijk$, and the 3-simplex $(i, j, k, l)$ by $ijkl$, in the event that the unit cell is a tetrahedron; the corresponding expressions for the cubic case are analogous.

The (abstract) *boundary* of a 1-simplex $ij$ is the set of its vertices:

$$\partial_1 \left( ij \right) = \{i, j\}. \tag{2.5}$$

The boundary of a 2-simplex $ijk$ consists of all of its consecutive vertex pairs – i.e., edges – and including the last letter followed by the first one:

$$\partial_2 \left( ijk \right) = \{\{i, j\}, \{j, k\}, \{k, i\}\}. \tag{2.6}$$

One can proceed analogously in each dimension and define the boundary $\partial_k$ of each $k$-simplex.

If one regards all permutations of letters in a word that represents a simplex as equivalent then one can think of that simplex as being *unoriented*. Hence, one can then think of a choice of permutation for the sequence of letters for a $k$-simplex as defining an *orientation* for that $k$-simplex. More generally, one can think of any permutation of the letters of $\sigma_a = ij\ldots k$ as being one or the other orientation for $\sigma_a$ according to the sign of the permutation (+ = even, − = odd). One arbitrarily assigns a + or − sign to $\sigma_a$ according to which type of permutation of $ij\ldots k$ defines it. That would make:

$$\partial_1 \left( ij \right) = (i, j) = - (j, i), \qquad \partial_2 \left( ijk \right) = (ij, jk, ki) = - (ji, jk, ki) = - (ij, kj, ki) = \ldots \tag{2.7}$$

The action of the boundary operator $\partial_k$ on $k$-simplexes extends to an action on $k$-chains with coefficients in $R$. If a $k$-chain takes the general form in (2.1) then its boundary will be:

$$\partial_k c_k = \sum_{i=1}^{r_k} a(i) \partial \sigma_k(i) = \sum_{j=1}^{r_{k-1}} \left[ \sum_{i=1}^{r_k} a(i) \partial^i_j \right] \sigma_{k-1}(j) , \tag{2.8}$$

which will be a $(k-1)$-chain such that the coefficient of the simplex $\sigma_{k-1}$ $(j)$ will be:

$$[\partial c_k](j) = \sum_{i=1}^{r_k} a(i) \partial^i_j . \tag{2.9}$$

The matrix $\partial_j^i$ is called the *incidence matrix* in dimension $k$, and its elements will be $-1$, $+1$, or $0$ depending upon whether the $(k-1)$-simplex $\sigma_{k-1}$ ($j$) belongs to the boundary of the $k$-simplex $\sigma_k$ ($i$) with one orientation or the other (in the first two cases) or it does not belong to the boundary, in the last case.

Hence, one can think of $\partial_k$ as a linear map $\partial_k : C_k$ ($\mathcal{S}_n$ ; $R$) $\rightarrow C_{k-1}$ ($\mathcal{S}_n$ ; $R$) . By definition, $\partial_0 : C_0$ ($\mathcal{S}_n$ ; $R$) $\rightarrow 0$ . A basic property of the boundary operator is that when it is applied twice in succession, that will always give zero:

$$\partial_{k-1}\,\partial_k = 0. \tag{2.10}$$

Hence: "The boundary of any boundary is zero."

The $R$-module $C_*$ ($\mathcal{S}_n$ ; $R$), together with the boundary operator $\partial : C_*$ ($\mathcal{S}_n$ ; $R$) $\rightarrow C_{*-1}(\mathcal{S}_n$ ; $R$), will then be called the *abstract chain complex with coefficients in R* for the abstract complex $\mathcal{S}_n$ .

A particularly useful class of 1-chains are the *paths*. Such a 1-chain takes the form:

$$c_1 = \sum_{i=1}^{p} \sigma_1(i)\,, \qquad \text{with} \quad \partial_1 \sigma_1\,(i) = \sigma_0\,(i+1) - \sigma_0\,(i)\,. \tag{2.11}$$

In other words, the endpoint of each 1-simplex is the initial point of the next one. That will make:

$$\partial c_1 = \sum_{i=1}^{p} \partial \sigma_1(i) = \sigma_0\,(p+1) - \sigma_0\,(1)\,. \tag{2.12}$$

$\partial c_1$ will vanish (i.e., $c_1$ will be a "1-cycle") iff $\sigma_0$ ($p+1$) = $\sigma_0$ (1) ; that is, the path must begin and end on the same point. Such a path will then be called a *loop*.

If any two vertices $\sigma_0$ ($i$) and $\sigma_0$ ($j$) can be joined by a path in a given complex then the complex will be called *path-connected*. Otherwise, the set of all $\sigma_0$ ($j$) that can be connected to $\sigma_0$ ($i$) will be called the *path component* of $\sigma_0$ ($i$). Since the relation "there exists a path between $\sigma_0$ ($i$) and $\sigma_0$ ($j$)" is reflexive, symmetric, and transitive, it is an equivalence relation, and the equivalence classes of that relation will be the path components.

*d. The augmented boundary operator.* – There is another variation on the definition of the boundary operator (see Greenberg [**13**]) that proves to be essential in discussing the equilibrium of forces in a mechanical network, as well as concepts such as the total charge or total mass of the network. It is the *augmented boundary operator* $\partial^{\#} : C_0$ ($\mathcal{S}_n$ ; $R$) $\rightarrow R$, and if a 0-chain takes the form:

$$c_0 = \sum_{i=1}^{r_0} m(i)\,\sigma_0(i) \tag{2.13}$$

then the definition of $\partial^{\#}$ will be:

$$\partial^{\#} c_0 \equiv \sum_{i=1}^{r_0} m(i) \,. \tag{2.14}$$

That is, $\partial^{\#} c_0$ is simply the sum of all the coefficients of the nodes.

One can show that ([1]):

$$\partial^{\#} \partial_1 = 0, \tag{2.15}$$

so that much has not changed from the usual boundary operator.

*e. The cone of a complex.* – One way to justify the use of the word "augmented" is to first add a 0-simplex, which we will denote by $\sigma_0$ (0), that is not contained in the generators of the complex $C_0(\mathcal{S}_n; R)$ .

One then adds all of the 1-simplexes $\{\sigma_1 (i, 0), i = 1, \ldots, r_0\}$ that connect each 0-simplex $\sigma_0(i)$ of the original complex to $\sigma_0$ (0) and then extend the boundary operator by:

$$\partial_1 \sigma_1 (i, 0) = \sigma_0 (0) - \sigma_0 (i), \quad \text{all } i. \tag{2.16}$$

One then gives the coefficient $m(i)$ to each 1-simplex $\sigma_1 (i, 0)$, where a 0-chain of the original complex is represented by $c_0$ as in (2.13). One can then define:

$$c_1^{\#} = \sum_{i=1}^{r_0} m(i) \sigma_1(i,0) \,. \tag{2.17}$$

One will then see that its boundary will be:

$$\partial c_1^{\#} = -\sum_{i=1}^{r_0} m(i) \sigma_0(i) + \left[\sum_{i=1}^{r_0} m(i)\right] \sigma_0(0) = -c_0 + (\partial^{\#} c_0) \,\sigma_0(0) \,. \tag{2.18}$$

Hence, we can extend the chains $c_0$ and $c_1$ to:

$$\bar{c}_0 = c_0 + \left[\sum_{i=1}^{r_0} m(i)\right] \sigma_0(0), \quad \bar{c}_1 = c_0 + c_1^{\#}, \tag{2.19}$$

resp., and if we originally had $\partial_1 c_1 = c_0$ then we will now have:

$$\partial_1 \bar{c}_1 = \left[\sum_{i=1}^{r_0} m(i)\right] \sigma_0(0) = (\partial^{\#} c_0) \,\sigma_0(0) \,. \tag{2.20}$$

---

([1])  The action of $\partial_1$ on a 1-chain will associate each $n(a) \sigma_1(a)$ with $n(a)$ at one boundary 0-simplex and $- n(a)$ at the other. Hence, upon summation each such pair will sum to zero.

One can turn the 1-simplexes of $C_1(\mathcal{S}_n; R)$ into 2-simplexes by joining $\sigma_0(0)$ to each of them, and so on up to dimension $n$, which will then produce $(n + 1)$-dimensional simplexes, but we shall be solely concerned with one-dimensional complexes in what follows, so we shall not dwell upon that fact.

The one-dimensional complex $\mathcal{C}[C_*(\mathcal{S}_1; R)]$ whose generators are $\{\sigma_0(0), \sigma_1(i, 0), i = 1, \ldots, r_0,\} \cup \{\sigma_0(i), \sigma_1(a), i = 1, \ldots, r_0, a = 1, \ldots, r_1\}$, which we shall then call the *cone of* $C_*(\mathcal{S}_1; R)$, will play a recurring role in what follows ([1]). The set of 1-simplexes $\{\sigma_1(i, 0), i = 1, \ldots, r_0,\}$ is referred to as the *star* of the 0-simplex $\sigma_0(0)$. (This definition of the star of a vertex can be extended to all $k$-simplexes that are incident with the given vertex for each dimension $k$ for higher-dimensional complexes.)

*f. Homology modules.* Since $\partial_k$ is linear, its kernel (i.e., the set of all $k$-chains with boundary zero) is also an $R$-module that is a submodule of $C_k(\mathcal{S}_n; R)$, and we denote this kernel by $Z_k(\mathcal{S}_n; R)$; an element of this $R$-module is called a *$k$-cycle*. Similarly, the image of $\partial_k$ will also be an $R$-submodule of $C_{k-1}(\mathcal{S}_n; R)$ that we denote by $B_{k-1}(\mathcal{S}_n; R)$; these $(k-1)$-chains are called $(k-1)$-*boundaries*. Note that since there are no $(n+1)$-chains (except 0), $B_n(\mathcal{S}_n; R) = 0$, while all 0-chains are cycles, so $Z_0(\mathcal{S}_n; R) = C_0(\mathcal{S}_n; R)$.

From (4.20), one sees that every $k$-boundary is also a $k$-cycle, so $B_k(\mathcal{S}_n; R)$ is an $R$-submodule of $Z_k(\mathcal{S}_n; R)$. The crucial question at the root of homology is whether the converse statement is true. In general, there might be $k$-cycles that do not bound $(k+1)$-chains. One can think of a triangle or square minus its interior points as examples of 1-cycles (the sequence of oriented edges) that do not bound a 2-chain. Hence, a non-bounding $k$-cycle represents a sort of "$(k+1)$-dimensional hole" in the space described by a chain complex.

One then defines the quotient module $H_k(\mathcal{S}_n; R) = Z_k(\mathcal{S}_n; R) / B_k(\mathcal{S}_n; R)$ to be the *homology module in dimension $k$* for the abstract simplicial complex $\mathcal{S}_n$. That is, an element of $H_k(\mathcal{S}_n; R)$ is an equivalence class of $k$-cycles under homology: Two $k$-cycles $z_k$ and $z_k'$ are said to be *homologous* iff their difference is the boundary of a $k+1$-chain:

$$z_k \sim z_k' \qquad \text{iff} \qquad z_k - z_k' = \partial c_{k+1} \qquad \text{for some } c_{k+1}. \qquad (2.21)$$

One then sees that when a $k$-cycle is a boundary it is "homologous to zero."

In particular, two vertices $v_1$ and $v_2$ are homologous iff there is some path from one to the other. Hence, the zero-dimensional homology module $H_0(\mathcal{S}_n; R)$ has a basis that is defined by the path-connected components of $\mathcal{S}_n$, so if $\mathcal{S}_n$ is path-connected then $H_0(\mathcal{S}_n;$

---

([1]) Actually, the more traditional definition of the cone of a one-dimensional complex (e.g., Hilton and Wylie [**11**]) would extend the 1-simplexes to 2-simplexes and produce a two-dimensional complex, so what we are calling the cone of the complex is really the 1-skeleton of the usual cone.

$R$) will be isomorphic to $R$; i.e., it will be a free $R$-module with one generator. Otherwise, if it has $p$ path components then it will be a free $R$-module with $p$ generators.

In the case of augmented homology, the 0-dimensional homology module $H_0^\#(\mathcal{S}_n; R)$, which consists of equivalence classes of augmented 0-cycles (so $\sum_{i=1}^{r_0} m(i) = 0$) that differ by a 0-boundary (under $\partial$), will be 0 when the complex is path-connected and have $p - 1$ generators, otherwise (see Greenberg [**13**]).

Once again, since there are no $(n+1)$-chains but 0, one will always have $H_n(\mathcal{S}_n; R) = Z_n(\mathcal{S}_n; R)$. A basis for $H_n(\mathcal{S}_n; R)$ will then consist of a minimal set of $n$-cycles, which a typical $n$-cycles will consist of a linear combination of minimal $n$-cycles with coefficients in $R$. In particular, we shall be mostly concerned with the case of $n = 1$.

It is important to note that whereas the module $C_k(\mathcal{S}_n; R)$ is free, the same thing will not be true for $H_k(\mathcal{S}_n; R)$, in general. It will generally be the direct sum of a free $R$-module $R^{b_k}$, where $b_k$ is called the $k^{th}$ *Betti number* of $\mathcal{S}_n$, and a finite number of finite cyclic groups, which one refers to as the *torsion part* of $H_k(\mathcal{S}_n; R)$. That means that for some $k$-cycles $z_k$ there will be a non-zero integer $N$ such that $N z_k = \partial c_{k+1}$ for some $c_{k+1}$. However, for the applications that we have in mind, the homology groups will typically be free and finitely-generated.

Generally, the Betti number in each dimension is equal to the number of "holes" of that dimension (plus one). For instance, one can say that an $n$-sphere has an $(n+1)$-dimensional hole in it, unless the interior points are included, in which case, it becomes contractible to a point or homologous to 0.

When one forms the alternating sum of the Betti numbers, the resulting integer is called the *Euler-Poincaré characteristic* of the abstract chain complex in question:

$$\chi[\mathcal{S}_n] = \sum_{k=0}^{n} (-1)^k b_k .\tag{2.22}$$

Interestingly, this number also equals the alternating sum $\sum_{k=0}^{n} (-1)^k r_k$ of the ranks of the modules $C_k(\mathcal{S}_n; R)$. For instance, in the case of a triangulated compact surface $\Sigma$:

$$\chi[\Sigma] = \# \text{ vertices} - \# \text{ branches} + \# \text{ faces},\tag{2.23}$$

which is the form that Euler himself used in his treatment of the Königsberg Bridges problem.

One can also define the *augmented homology module* $H_0^\#(\mathcal{S}_n; R)$ to be the quotient of the $R$-module $Z_0^\#(\mathcal{S}_n; R)$ of augmented 0-cycles $z_0^\#$, for which $\partial^\# z_0^\# = 0$, by the $R$-module $B_0(\mathcal{S}_n; R)$ of ordinary 0-boundaries. If $\mathcal{S}_n$ is path-connected then $H_0^\#(\mathcal{S}_n; R) = 0$, but if $\mathcal{S}_n$ has $p$ path components then $H_0^\#(\mathcal{S}_n; R)$ will be a free $R$-module with $p - 1$ generators (see Greenberg [**13**]).

*g. Chain maps.* – When we get into motions and deformations of mechanical networks, it will be important to have the concept of a "chain map" in hand. Basically, such a map takes chain complexes to chain complexes in such that way that boundaries will go to boundaries. Hence, if $\mathcal{S}_n$ and $\overline{\mathcal{S}}_n$ are two abstract simplicial complexes then for each $0 \leq k \leq n$, one will have a linear map $f_k : C_k(\mathcal{S}_n ; R) \rightarrow C_k(\overline{\mathcal{S}}_n ; R)$ that takes the generating $k$-simplexes $\sigma_k$ of $C_k(\mathcal{S}_n ; R)$ to generating $k$-simplexes of $C_k(\overline{\mathcal{S}}_n ; R)$, but not necessarily in a bijective sort of way. The condition that those maps $f_k$ must preserve boundaries takes the form:

$$\partial_{k-1} \cdot f_k = f_{k-1} \cdot \partial_k ; \tag{2.24}$$

i.e., when this is evaluated on any $k$-chain $c_k$ :

$$\partial_{k-1}(f_k(c_k)) = f_{k-1}(\partial_k c_k) . \tag{2.25}$$

As a result, a chain map will take cycles to cycles and boundaries to boundaries. That means that a chain map will "descend to homology," in the sense that homologous cycles with go to homologous cycles. That is, if $z'_k - z_k = \partial_{k+1} c_{k+1}$ is a pair of homologous $k$-cycles and $f_k$ is a chain map then:

$$f_k(z'_k) - f_k(z_k) = f_k(z'_k - z_k) = f_k(\partial_{k+1} c_{k+1}) = \partial_{k+1}(f_{k+1}(c_{k+1})) ;$$

i.e.:

$$f_k(z'_k) - f_k(z_k) = \partial_{k+1}(f_{k+1}(c_{k+1})) . \tag{2.26}$$

Hence, the images of $z_k$ and $z'_k$ under $f_k$ will also be homologous.

That means that each map $f_k$ will define a corresponding linear map $\overline{f}_k : H_k(\mathcal{S}_n ; R) \rightarrow H_k(\overline{\mathcal{S}}_n ; R)$ that takes homology classes of $k$-cycles to homology classes of $k$-cycles. Of particular interest are the chain maps that induce isomorphisms in each dimension. When a chain isomorphism exists, the chain complexes are said to be *homologically equivalent*.

*h. Cohomology.* – Since modules are a generalization of vector spaces where the scalars belong to a ring that does not have to form a field, one can examine all of the usual constructions of linear algebra to see if they generalize, as well. One of those constructions that can be generalized is the concept of the "dual space $V^*$" to a given vector space $V$. An element $\phi \in V^*$ is defined (in linear algebra) to be a linear functional on $V$, so for every pair of vectors $\mathbf{v}, \mathbf{w} \in V$ and every pair of scalars $\alpha, \beta \in \mathbb{R}$, one will have:

$$\phi(\alpha \mathbf{v} + \beta \mathbf{w}) = \alpha \phi(\mathbf{v}) + \beta \phi(\mathbf{w}) . \tag{2.27}$$

The generalization from a vector space to a $R$-module is straightforward: One simply defines linear functionals on $C_k(\mathcal{S}_n ; R)$ to be maps $c^k : C_k(\mathcal{S}_n ; R) \rightarrow R$ such that for every pair of $k$-chains $c_k , c'_k \in C_k(\mathcal{S}_n ; R)$ and every pair of scalars $\alpha, \beta \in R$ one will have:

$$c^k \left( \alpha \, c_k + \beta \, c'_k \right) = \alpha \, c^k \left( c_k \right) + \beta \, c^k \left( c'_k \right) . \tag{2.28}$$

In particular, $c^k \left( c_k \right)$ will always be an element of $R$. Hence, one says that these linear functionals, which one calls *k-cochains on* $\mathcal{S}_n$, "have values in $R$." We shall denote the $R$-module of linear functionals on $C_k \left( \mathcal{S}_n ; R \right)$ by $C^k \left( \mathcal{S}_n ; R \right)$ for each $k$ and refer to it as the *R-module of k-cochains on* $\mathcal{S}_n$.

As long as one is concerned with finitely-generated free $R$-modules, one can associate any set of generators $\{ \sigma_k \left( i \right), \ i = 1, \ \ldots, \ r_k \}$ for $C_k \left( \mathcal{S}_n ; R \right)$ with a "reciprocal" set of generators $\{ \sigma^k \left( i \right), \ i = 1, \ \ldots, \ r \}$ for $C^k \left( \mathcal{S}_n ; R \right)$ in the same way that one associates a reciprocal basis for a vector space with a basis. Namely, the reciprocal generators have the defining property that:

$$< \sigma^k \left( i \right), \sigma_k \left( j \right) > = \delta_{ij} = \left\{ \begin{array}{ll} 0 & i \neq j \\ 1 & i = j \end{array} \right. \quad \text{for all } i, j, \tag{2.29}$$

in which the bracket refers to the canonical bilinear pairing of elements of a dual space with elements of the space by evaluating a linear functional on an element.

An arbitrary $k$-cochain can then be represented as a formal sum:

$$c^k = \sum_{i=1}^{r_k} n \left( i \right) \sigma^k \left( i \right), \qquad n \left( i \right) \in R, \tag{2.30}$$

and its evaluation on a $k$-chain $c_k$ of the form (4.14) will yield:

$$< c^k, c_k > = \sum_{i=1}^{r_k} \sum_{j=1}^{r_k} n(i) \, m(j) < \sigma^k(i), \sigma_k(j) > = \sum_{i=1}^{r_k} \sum_{j=1}^{r_k} n(i) \, m(j) \, \delta_j^i = \sum_{i=1}^{r_k} n(i) \, m(i) . \tag{2.31}$$

The boundary operator $\partial_k : C_k \left( \mathcal{S}_n ; R \right) \rightarrow C_{k-1} \left( \mathcal{S}_n ; R \right)$ can be "pulled back" to define a *coboundary operator* $\delta^{k-1} : C^{k-1} \left( \mathcal{S}_n ; R \right) \rightarrow C^k \left( \mathcal{S}_n ; R \right)$, so its definition will take the simple form:

$$< \delta^{k-1} \, c^{k-1}, c_k > = < c^{k-1}, \partial_k \, c_k > . \tag{2.32}$$

It shares the basic property of $\partial_k$ that its "square" vanishes:

$$\delta^k \, \delta^{k-1} = 0, \tag{2.33}$$

which will follow immediately from the definition (2.32). Hence, the coboundary of a coboundary is zero.

If $c^k$ is a $k$-cochain of the form (2.30) then its coboundary can be expressed in the form:

$$\delta^k \, c^k = \sum_{i=1}^{r_k} n \left( i \right) \delta^k \sigma^k \left( i \right) = \sum_{j=1}^{r_{k+1}} \left[ \sum_{i=1}^{r_k} n \left( i \right) [\delta]_i^j \right] \sigma^{k+1} \left( j \right) . \tag{2.34}$$

The matrix $[\delta]_i^j$ proves to be simply the transpose of the corresponding incidence matrix $\partial_i^j$ :

$$[\delta]_i^j = \partial_j^i.  \qquad (2.35)$$

One can define an "augmented" coboundary $\delta^{\#} : R \to C^0(\mathcal{S}_n ; R)$, $a \mapsto \delta^{\#}a$, which would be the transpose of $\partial^{\#}$. If one evaluates $\delta^{\#}a$ on any 0-chain $c_0$ then one will get:

$$<\delta^{\#}a, c_0> = <a, \partial^{\#}c_0> = a \, \partial^{\#}c_0 . \qquad (2.36)$$

The definition of $\delta^k$ allows one to define analogues to $Z_k$ , $B_k$ , and $H_k$ in the form the $R$-module $Z^k$ of of $k$-cocyles , the $R$-module $B^k$ of of $k$-coboundaries , and the quotient $R$-module $H^k = Z^k / B^k$ of $k$-cocycles that differ by a $(k–1)$-coboundary; i.e., *cohomologous $k$-cocyles.* Hence, $z^k$ is a $k$-cocyle iff $\delta^k z^k = 0$, $b^k$ is a $k$-coboundary iff $b^k = \delta^{k-1} c^{k-1}$ for some $(k-1)$-cochain $c^{k-1}$, and two $k$-cocycles $z^k$, $z'^k$ are cohomologous iff there is some $c^{k-1}$ such that:

$$z'^k - z^k = \delta^{k-1} c^{k-1} . \qquad (2.37)$$

Although the $R$-modules $Z^k$ and $B^k$ are sub-modules of $C^k$, and therefore dual spaces to $Z_k$ , $B_k$ , nonetheless, the resulting cohomology module $H^k$ does not have to be dual to $H_k$. However, as long as $H_k$ is free that will be the case.

If one has a chain map $f_* : C_*(\mathcal{S}_n ; R) \to C_*(\bar{\mathcal{S}}_n ; R)$, and therefore a linear map $f_k : C_k(\mathcal{S}_n ; R) \to C_k(\bar{\mathcal{S}}_n ; R)$ for each $k$ that commutes with the boundary operator, then one can define its transpose map $f^* : C^*(\bar{\mathcal{S}}_n ; R) \to C^*(\mathcal{S}_n ; R)$, by "pulling back" a $k$-cochain $\bar{c}^k \in C^k(\bar{\mathcal{S}}_n ; R)$ along $f^k$ to a $k$-cochain $f^k \bar{c}^k \in C^k(\mathcal{S}_n ; R)$ ; that is, if $c_k$ is a $k$-chain in $C_k(\mathcal{S}_n ; R)$ then the evaluation of $f^* \bar{c}^k$ on $c_k$ will give:

$$< f^k \bar{c}^k, c_k > = <\bar{c}^k, f_k(c_k) > . \qquad (2.38)$$

The fact that $f_*$ commutes with the boundary operator $\partial$ translates into the fact that $f^*$ commutes with the boundary operator $\delta$ :

$$f^* \cdot \delta = \delta \cdot f^* . \qquad (2.39)$$

Hence, cocycles will pull back to cocycles, coboundaries will pull back to coboundaries, and cohomologous cocycles will pull back to cohomologous cocycles. As a result, the pull back map $f^*$ will induce a linear map $\bar{f}^* : H^*(\bar{\mathcal{S}}_n ; R) \to H^*(\mathcal{S}_n ; R)$ .

## 3. Electrical networks.

– An electrical network defines an elementary example of a one-dimensional abstract simplicial complex with a finite number of generators in both dimensions. Indeed, until one starts going into finer details about the circuit components (resistor, capacitors, inductors, etc.), one typically does not need to even have a geometric realization of it in some geometric space. Typically, the references that deal with the topology of networks [**14**] prefer to model the network as a graph, but the lectures of Lefschetz [**9**] are an exception to that. One can see that when it becomes necessary to look at the circuit components in more detail, the dimension of the complex will generally increase, since one has to contend with the areas of capacitor plates, as well as the electric fields between them, and the magnetic fields of inductors, among other things.

*a. The basic complex of a network.* – An electrical network $\mathcal{S}_1 = \{N, B, \partial_a^i, R\}$ consists of two finite sets, namely, the set $N = \{\sigma_0(i)\ i = 1, \ldots, r_0\}$ of *nodes*, at which components connect to each other, and the set $B = \{\sigma_1(a), a = 1, \ldots, r_1\}$ of *branches*, which connect nodes to each other and often contain circuit components, and an incidence matrix $\partial_a^i,\ i = 1, \ldots, r_0,\ a = 1, \ldots, r_1$, which represents the matrix of the boundary operator $\partial_1 : C_1(\mathcal{S}_1\ ;\ R) \to C_0(\mathcal{S}_1\ ;\ R)$ that says which oriented pairs of nodes are the endpoints of each branch.

The symbol $R$ represents the coefficient ring, which will consist of *signals*. There are two basic types of signals that one deals with in electrical networks: ones that are constant in time and ones that are not. The former are referred to as "direct currents (DC)," while the latter *include* the "alternating currents (AC)." However, there are typically more general time-varying currents at work than the basic sinusoidal kind, which serve as basis vectors in an infinite-dimensional vector space of periodic functions by way of Fourier series expansions. Hence, we shall recognize two basic coefficient rings to be our rings of signals: DC signals, which belong to the field $\mathbb{R}$, and time-varying signals, which belong to the ring $C^1(\mathbb{R})$ of differentiable real-valued functions $f(t)$ of time $t \in \mathbb{R}$.

The generators for both $R$-modules $C_0(\mathcal{S}_1\ ;\ R)$ and $C_1(\mathcal{S}_1\ ;\ R)$ are defined by the sets $N$ and $B$. Hence, a 0-chain $c_0 \in C_0(\mathcal{S}_1\ ;\ R)$ and a 1-chain $c_1 \in C_1(\mathcal{S}_1\ ;\ R)$ will take the forms:

$$c_0 = \sum_{i=1}^{r_0} m(i)\,\sigma_0(i)\,, \qquad c_1 = \sum_{a=1}^{r_1} n(a)\,\sigma_1(a)\,, \tag{3.1}$$

respectively, in which all $m(i)$ and $n(a)$ are elements of $R$.

The only non-trivial boundary operator is the one that acts upon 1-chains, and its matrix with respect to the chosen generators is $\partial_a^i$, so the boundary of an arbitrary 1-chain $c_1$ will be:

$$\partial c_1 = \sum_{a=1}^{r_1} n(a)\,\partial\sigma_1(a) = \sum_{i=1}^{r_0}\left[\sum_{a=1}^{r_1} n(a)\,\partial_a^i\right]\sigma_0(i) = \sum_{i=1}^{r_0} m(i)\,\sigma_0(i)\,, \tag{3.2}$$

in which we have defined the coefficients $m(i)$ of the 0-simplexes of the form $\partial c_1$ by:

$$m\,(i) = \sum_{a=1}^{r_1} n\,(a)\,\partial_a^i \; . \qquad (3.3)$$

Hence, the coefficient $m\,(i)$ for a given node $i$ will be equal to a sum of the (signed) coefficients of the branches that are incident at that node.

*b. Kirchhoff's law of currents.* – The last statement sounds suspiciously reminiscent of Kirchhoff 's law of currents, namely, that the sum of the currents at a node is equal to zero, or more generally, to the time derivative of the charge that is stored at the node. In order to justify that observation, we will associate the nodes with time-varying charges $\{Q\,(i),\, i = 1,\, \ldots,\, r_0\}$ and currents $\{I\,(a),\, a = 1,\, \ldots,\, r_1\}$ with the branches and change the generic definitions in (3.1) to take the forms:

$$Q = \sum_{i=1}^{r_0} Q\,(i)\,\sigma_0\,(i), \qquad I = \sum_{a=1}^{r_1} I\,(a)\,\sigma_1\,(a), \qquad (3.4)$$

respectively. We shall call the 0-chain $Q$ the *charge distribution* in the circuit and the 1-chain $I$ the *current distribution* in it.

We define the action of the time derivative operator on the 0-chain $Q$ by its action on the coefficients:

$$\dot{Q} \equiv \frac{dQ}{dt} = \sum_{i=1}^{r_0} \dot{Q}\,(i)\,\sigma_0\,(i), \qquad \dot{Q}\,(i) = \frac{dQ(i)}{dt}, \qquad \text{for each } i \, . \qquad (3.5)$$

If we then examine the meaning of the equation:

$$\partial I = \frac{dQ}{dt} \qquad (3.6)$$

then we will see that since:

$$\partial I = \sum_{a=1}^{r_1} I\,(a)\,\partial\sigma_1\,(a) = \sum_{i=1}^{r_0} \left[ \sum_{a=1}^{r_1} I\,(a)\,\partial_a^i \right] \sigma_0\,(i) \, , \qquad (3.7)$$

we will have:

$$\dot{Q}\,(i) = \sum_{a=1}^{r_1} I\,(a)\,\partial_a^i \, , \qquad (3.8)$$

which says that the time rate of change of the charge that is stored at the node $i$ equals the signed sum of the currents in the branches that are incident at that node. We have then duplicated a generalization of Kirchhoff's laws of currents; it is also a statement of the *balance of charge* in the system. The usual law of currents is obtained as a corollary when one imposes the *conservation of charge*, which would make $dQ\,(i)\,/\,dt$ vanish at each node. One can then phrase the law as:

**Kirchhoff's law of currents:**

*Charge is conserved in a circuit iff the current distribution takes the form of a 1-cycle:*

$$\partial I = 0. \tag{3.9}$$

Note that when we go to the augmented boundary operator, we will get:

$$\partial^{\#} Q = \sum_{i=1}^{r_1} Q(i), \tag{3.10}$$

which is simply the sum of the charges at the nodes.

We can then differentiate this and get:

$$\partial^{\#} \dot{Q} = \sum_{i=1}^{r_1} \dot{Q}(i), \tag{3.11}$$

which is the time rate of change of the total charge. When (3.6) is applicable this will imply that the time rate of change of the charge 0-chain must be an augmented 0-cycle.

We can also extend the original network with a 0-simplex $\sigma_0(0)$ and the 1-simplexes $\sigma_1(0, i)$ that connect $\sigma_0(0)$ to each node $i$, such that:

$$\partial \sigma_1(0, i) = \sigma_0(0) - \sigma_0(i) \qquad \text{for all } i. \tag{3.12}$$

If the current associated with each $\sigma_1(i, 0)$ is $\dot{Q}(i)$ then the boundary of the extended current 1-chain:

$$\bar{I} = I + \sum_{i=1}^{r_0} \dot{Q}(i)\, \sigma_1(0, i) \tag{3.13}$$

will be:

$$\partial \bar{I} = \partial I + \left[ \sum_{i=1}^{r_0} \dot{Q}(i) \right] \sigma_0(0) - \sum_{i=1}^{r_0} \dot{Q}(i)\, \sigma_0(i) = \left[ \sum_{i=1}^{r_0} \dot{Q}(i) \right] \sigma_0(0) = \partial^{\#} \dot{Q}\, \sigma_0(0), \tag{3.14}$$

when (3.6) applies. Hence, $\partial \bar{I}$ simply assigns the time derivative of the total charge to the 0-simplex $\sigma_0(0)$. That will vanish iff the total charge is constant, so:

**Theorem (conservation of total charge):**

*The following are equivalent:*

*i)  The total charge that is associated with the nodes of an electrical network is constant in time.*

**ii)** *The 0-chain of time derivatives of charge is an augmented 0-cycle.*

**iii)** *The extended current 1-chain is a 1-cycle.*

*c. Kirchhoff's law of voltages.* – It is quite remarkable that all one needs to do to get to get Kirchhoff's law of voltages (namely, that the sum of the voltage drops around any loop composed of branches is zero) is to look at the corresponding dual picture in terms of the cohomology of the network with values in the ring of signals.

One starts by noting that a 1-cochain is a coboundary iff it gives zero whenever it is evaluated on a 1-cycle, e. g., a loop composed of branches. This follows directly from the definition of the coboundary:

$$< \delta c^0, z_1 > = < c^0, \partial z_1 > = 0. \tag{3.15}$$

Hence, one defines the *voltage distribution* in the circuit to be the 0-cochain $V$ that has its values in $R$, so the coefficients of $V$ will be $\{V_i, i = 1, \ldots, r_0\}$:

$$V = \sum_{i=1}^{r_0} V(i)\,\sigma^0(i)\,. \tag{3.16}$$

The coboundary of $V$ will then be:

$$\delta V = \sum_{i=1}^{r_0} V(i)\,\delta\sigma^0(i) = \sum_{a=1}^{r_1} \Delta V(a)\,\sigma^1(a)\,, \qquad \Delta V(a) = \sum_{i=1}^{r_0} V(a)\,\partial_a^i\,. \tag{3.17}$$

Hence, the coefficient $\Delta V(a)$ represents the "voltage drop" across the branch $\sigma_1(a)$. Namely, if $\partial\sigma_1(a) = \sigma_0(i) - \sigma_0(j)$ then:

$$\Delta V(a) \equiv <\delta V,\,\sigma_1(a)> = <V,\,\partial\sigma_1(a)> = V(i) - V(j)\,. \tag{3.18}$$

When one evaluates $\delta V$ on any 1-cycle, one will get 0, from (3.15). One can generalize the definition of $\delta V$ in (3.17) to simply a 1-cochain with values in $R$:

$$\Delta V \equiv \sum_{a=1}^{r_1} \Delta V(a)\,\sigma^1(a)\,. \tag{3.19}$$

Let $z_1$ be a 1-cycle:

$$z_1 = \sum_{a=1}^{r_1} m(a)\,\sigma_1(a)\,. \tag{3.20}$$

so

$$0 = \partial z_1 = \sum_{i=1}^{r_0} \left[ \sum_{a=1}^{r_1} n(a)\,\partial_a^i \right] \sigma_0(i)\,, \tag{3.21}$$

which says that:

$$\sum_{a=1}^{r_1} m(a)\,\partial_a^i = 0 \qquad \text{for every } i\,. \tag{3.22}$$

When one evaluates $\Delta V$ on $z_1$ that will yield:

$$<\Delta V, z_1> = <\sum_{a=1}^{r_1} \Delta V(a)\,\sigma^1(a),\ \sum_{b=1}^{r_1} n(b)\,\sigma_1(b)> = \sum_{a=1}^{r_1} \sum_{b=1}^{r_1} \Delta V(a)\,m(b)\,\delta_b^a$$

$$= \sum_{a=1}^{r_1} m(a)\,\Delta V(a)\ .$$

This is basically the sum of the voltage drops around the branches of the cycle $z_1$. Hence, if we represent $<\Delta V, z_1>$ by the signal $dE / dt\ (z_1)$ where $E$ stands for the energy that is being add to or taken from the cycle $z_1$ then the definition:

$$\frac{dE}{dt}(z_1) = <\Delta V, z_1> = <V, \partial z_1> = 0 \qquad (3.23)$$

represents the balance of energy in that 1-cycle, which gives:

**Kirchhoff's law of voltages:**

*Energy is conserved in a circuit iff the voltage drop distribution is a 1-coboundary:*

$$\Delta V = \delta V. \qquad (3.24)$$

More generally, if one defines the *power* 1-cochain by:

$$\frac{dE}{dt} = \sum_{a=1}^{r_1} \dot{E}(a)\,\sigma^1(a) \qquad (3.25)$$

then one will have:

**Theorem (balance of energy):**

*For any electrical network, one will have:*

$$\frac{dE}{dt} = \Delta V. \qquad (3.26)$$

Hence:

**Corollary:**

*The power 1-cochain is a coboundary.*

**4. Geometric realizations of abstract simplicial complexes.** – The abstract simplicial complexes that we shall be concerned with in the study of mechanical networks will be one-dimensional complexes for which the nodes or vertices will be massive points in space at which external forces act, and the branches that connect them might take the form of material links that are associated with the displacements that take one vertex to the other and the forces that act inside of them. Hence, it will be useful to represent the abstract objects of $\mathcal{S}_n$ in terms of geometric objects in the space that those points inhabit; i.e., we will need to define a *geometric realization* of the complex. For our present purposes, it will be sufficient to let that space be an $n$-dimensional affine space $A^n$, where $n$ will generally be less than four, so we shall begin with a brief review of the basic definitions that concern affine spaces

*a. Affine spaces.* − An *affine space* is a set $A^n$ whose elements are thought of as points and an action of the translation group $\mathbb{R}^n$ on $A^n$:

$$\mathbb{R}^n \times A^n \to A^n, (\mathbf{s}, x) \mapsto x + \mathbf{s}.$$

Note that the usage of the "+" sign in this context is purely symbolic, since one does not generally assume that an affine space has any underlying notion of addition, as a vector space would.

There are two basic properties of the action:

*i*)   The action of 0 is trivial:

$$x + 0 = x. \tag{4.1}$$

*ii*)   When one composes a translation by $\mathbf{a}$ with a further translation by $\mathbf{b}$ the result will be:

$$(x + \mathbf{a}) + \mathbf{b} = x + (\mathbf{a} + \mathbf{b}). \tag{4.2}$$

Thus, the net effect of the succession of translations is a single translation by $\mathbf{a} + \mathbf{b}$. Naturally, that extends by recursion to a sequence of any finite number of translations. Since the translation group $\mathbb{R}^n$ is Abelian, it is not necessary to specify the sequence uniquely.

If we define the notation:

$$x - \mathbf{s} \equiv x + (-\mathbf{s}) \tag{4.3}$$

then that will imply:

$$(x + \mathbf{s}) - \mathbf{s} = x + (\mathbf{s} - \mathbf{s}) = x. \tag{4.4}$$

The most immediate example of an affine space is $\mathbb{R}^n$ itself. The action $\mathbb{R}^n$ on itself is just vector addition, which is commutative, so there is not need to distinguish between a left action and right action.

An element $\mathbf{s}\,(x,\,y) \in \mathbb{R}^n$ is called a *displacement vector* when its action can be expressed as an equation by saying that $\mathbf{s}\,(x,\,y)$ takes the point $x$ to the point:

$$y = x + \mathbf{s}\,(x,\,y)\,. \tag{4.5}$$

If one re-expresses this equation as:

$$y - x = \mathbf{s}\,(x, y) \tag{4.6}$$

then one can also use (4.6) as the definition of $\mathbf{s}\,(x, y)$ ; that is, there is also an antisymmetric binary map:

$$\mathbf{s} : A^n \times A^n \to \mathbb{R}^n,\, (x, y) \mapsto \mathbf{s}\,(x, y) \tag{4.7}$$

that is equivalent to the original group action.

The fact that the latter map is antisymmetric, i.e.:

$$\mathbf{s}\,(y, x) = -\,\mathbf{s}\,(x, y)\,, \tag{4.8}$$

is simply the statement that the displacement vector that takes $y$ to $x$ is the negative of the displacement vector that takes $x$ to $y$.

A further restriction upon the action of $\mathbb{R}^n$ on any $n$-dimensional affine space is that it must be "simply transitive." Saying that the group action on $A^n$ is *transitive* is equivalent to saying that for every pair $(x, y)$ of points in $A^n$ there is *some* displacement vector $\mathbf{s} \in \mathbb{R}^n$ that will take $x$ to $y$; saying that it is *simply* transitive means that the displacement vector will be *unique* in every case. Hence, the map that is defined by (4.7) will be defined and single-valued for all pairs in $A^n \times A^n$; i.e., it is a *function*.

Whenever one chooses a point $O \in A^n$ that will serve as an arbitrary "origin" for the affine space, one can define a bijective map:

$$\mathbf{s}_O : A^n \to \mathbb{R}^n,\, (x) \mapsto \mathbf{s}\,(O, x) \tag{4.9}$$

that associates every point $x \in A^n$ with a unique vector $\mathbf{s}\,(O, x)$ in $\mathbb{R}^n$ that represents the displacement vector that will take the point $O$ to the point $x$. Because of this one-to-one correspondence, one then refers to $\mathbb{R}^n$ as the *vector space on which $A^n$ is modeled.* One can also regard $\mathbb{R}^n$ as $T_O A^n$, namely, the tangent space to $A^n$ at $O$.

When one chooses a different origin $O' \in A^n$, one can first define the vector:

$$\Delta O = O' - O, \tag{4.10}$$

and then a different map to $\mathbb{R}^n$:

$$\mathbf{s}_{O'} : A^n \to \mathbb{R}^n,\, (x) \mapsto \mathbf{s}\,(O', x)\,. \tag{4.11}$$

One can then relate $\mathbf{s}_{O'}$ to $\mathbf{s}_O$ quite simply by substituting (4.10) in (4.9) and keeping (4.2) in mind:

$$\mathbf{s}_{O'}(x) = \mathbf{s}\,(O + \Delta O, x) = x - (O + \Delta O) = (x - O) - \Delta O = \mathbf{s}_O\,(x) - \Delta O\,,$$

or, more briefly:

$$\mathbf{s}_{O'}(x) = \mathbf{s}_O(x) - \Delta O .\qquad(4.12)$$

*b. Geometric realization of abstract simplexes.* – From now on, we shall think of abstract simplexes as the indices for things that live in an *n*-dimensional affine space.

In order to realize an abstract complex $\mathcal{S}_m$ in $A^n$, one first associates the abstract 0-simplexes with distinct points in $A^n$ by way of a one-to-one map $\sigma_0 : S_0 \to A^n$, $i \mapsto \sigma_0(i)$. The points $\sigma_0(i)$ in $A^n$ will then be thought of as *geometric 0-simplexes* of the complex and the set of all of them will be denoted by $L_0$.

The geometric realization of $S_1$ then maps each index pair $(i, j)$ in $S_1$ to a distinct ordered line segment ([1]) $\sigma_1(i, j)$ in $A^n$ that connects the point $\sigma_0(i)$ to the $\sigma_0(j)$ in that order; hence, the line segment $\sigma_1(j, i)$ would be distinct from $\sigma_1(i, j)$. If the points $\sigma_0(i)$ that correspond to the $\sigma_0(j)$ are denoted by $x_0$ and $x_1$, respectively, and the displacement vector between them is $\mathbf{s} = x_1 - x_0$ then the points of the line segment $\sigma_1(i, j)$ that connects them can be defined by:

$$x(t) = x_0 + t\,\mathbf{s}, \qquad 0 \le t \le 1.\qquad(4.13)$$

The points in $A^n$ that realize the 1-simplex $\sigma_1(i, j)$ are called the *carrier* of that simplex. The points in $A^n$ that carry the line segments $\sigma_1(i, j)$ that correspond to vertex pairs $(i, j)$ in $S_1$ will then be the *geometric 1-simplexes* of the complex and we will denote that set by $L_1$.

We can continue to define geometric realizations of the higher-dimensional abstract simplexes in $A^n$, although even in two dimensions, we have a number of possible ways of defining 2-simplexes, such as triangular areas, quadrilateral ones, etc. Since we shall be concerned with only one-dimensional complexes in this article, we shall not go into such irrelevant complexities, however. However, for the moment, we shall assume that such higher-dimensional simplexes have been defined in some way.

One can give a definition of the orientation of a geometric simplex that corresponds to the definition for abstract simplexes. In the case of line segments, one can think of an orientation as defining a sense of motion from one vertex to the other, while for a triangle or parallelogram an orientation is a sense of rotation for a circuit around the vertices by way of the edges. One can also think of a 2-simplex as being oriented by a choice of direction for a normal line to it. Ultimately, an orientation for an abstract simplex should correspond to an orientation for its geometric realization.

Previously, we called $\mathcal{S}_p = S_0 \cup \ldots \cup S_p$ the *p*-skeleton of the abstract simplicial complex. We now define the corresponding *p-skeleton* $\Lambda_p$ of its geometric realization to be $\Lambda_k = L_0 \cup \ldots \cup L_k$. In either case, if the dimension of the complex is *m* then the entire complex will coincide with its *m*-skeleton. We shall similarly refer to $\Lambda_m$ as the "geometric complex."

---

([1]) As far as topology is concerned, the use of a straight line segment is not necessary, and one can just as well use a continuous non-self-intersecting curve.

*c.   Geometric k-chains.* – We shall first direct our attention to the geometric simplexes.  For this section, the dimension of the complex will be $m \leq n$ .

A *geometric k-chain* $c_k$ is again defined to be a finite formal sum over all of the oriented geometric $k$-simplexes $\sigma_k(i)$, $i = 1, \ldots, r_k \in L_k$ with coefficients $m(i)$ that belong to the ring $R$:

$$c_k = \sum_{i=1}^{r_k} m(i)\, \sigma_k(i)\,. \tag{4.14}$$

Once again, $m(i)$ might be 0 in some cases.

The set $C_k(\Lambda_m ; R)$ of all geometric $k$-chains formed from the $k$-simplexes of the geometric realization $L$ of $S$ in $A^n$, when it is given the scalar multiplication by integers (2.2) and the addition (2.3), will then define a free $R$-module that is generated by the set $L_k$ .

The rank of $C_k(\Lambda_m ; R)$ still equals the number $r_k$ of $k$-simplexes in $S_k$ .  The fact that we are dealing with finite simplicial complexes implies that each $C_k(\Lambda_m ; R)$ has a finite rank.

Since non-zero $k$-chains do not exist for $k > m$, one already has $C_k(\Lambda_m ; R) = 0$, $k > m$.

*d. The geometric boundary operator.*  One begins by defining the boundaries of the basic simplexes of the geometrical complex $\Lambda_m$ in each dimension.

By definition, a 0-simplex will have 0 for its boundary.  Hence:

$$\partial \sigma_0(i) = 0 \tag{4.15}$$

for all $i$ in $S_0$ ; from now on, we shall refer to all $\partial_k$ by the same symbol $\partial$, for brevity.

One can then extend this by linearity to all 0-chains as the constant map $\partial : C_0(\Lambda_m; R) \rightarrow 0$, which takes all 0-chains to zero.

The boundary of each oriented 1-simplex $\sigma_1(i, j)$ will be, by definition:

$$\partial \sigma_1(i, j) = \sigma_0(j) - \sigma_0(i), \tag{4.16}$$

which is then a 0-chain, not a 0-simplex.

The linear extension of this to all 1-chains is then $\partial : C_1(\Lambda_m ; R) \rightarrow C_0(\Lambda_m ; R)$, $c_1 \mapsto \partial c_1$, and if $c_1 = \sum_{a=1}^{r_1} n(a)\, \sigma_1(a)$ then:

$$\partial c_1 = \sum_{a=1}^{r_1} n(a)\, \partial \sigma_1(a) = \sum_{i=1}^{r_0} \left[ \sum_{a=1}^{r_1} n(a)\, \partial_a^i \right] \sigma_0(i)\,, \tag{4.17}$$

in which the matrix $\partial_a^i$ is the incidence matrix of the 1-simplexes, as before.  Its elements will be $-1, 0, +1$ according to whether $\sigma_0(i)$ is incident on the "initial" end of $\sigma_1(a)$, not incident on $\sigma_1(a)$, or incident on the "final" end of it, respectively.

If we let $m(i)$ denote the coefficient of $\sigma_0(i)$ in (4.17) then we can say that:

$$m(i) = \sum_{a=1}^{r_i} n(a) \partial_a^i \ . \tag{4.18}$$

One can then extend this definition of the boundary operator to higher-dimensional chains to give a linear operator $\partial : C_k(\Lambda_m ; R) \to C_{k-1}(\Lambda_m ; R)$ . That is, one first defines $\partial \sigma_k(i)$, and then extends to formal sums of $k$-simplexes by linearity. Thus, if $c_k$ takes the form that it has in (4.14), by allowing some $m(i)$ to be zero, then:

$$\partial c_k = \sum_{i=1}^{r_k} m(i) \partial \sigma_k(i) = \sum_{i=1}^{r_{k-1}} \left[ \sum_{j=1}^{r_k} m(j) \partial_i^j \right] \sigma_{k-1}(i) , \tag{4.19}$$

in which $r_{k-1}$ represents the total number of all $(k-1)$-simplexes.

Of course, in this article, we shall have no use for the higher-dimensional generalization, but an analogous construction will become useful when we go on to continuous systems of interacting masses in the second part of this series of articles.

As in the abstract case, a basic property of the boundary operator is that its "square" always vanishes:

$$\partial^2 = \partial_{k-1} \partial c_k = 0 \qquad\qquad [\text{i.e.,} \quad \partial_{k-1}(\partial c_k) = 0 \text{ for any } c_k]. \tag{4.20}$$

The definition of the boundary operator that we have given seems rather abstract, on first glance, since it starts with algebraic structures – namely, the various $C_k(\Lambda_m ; R)$ – whose structures as $R$-modules relate to only the number of elements that generate them, but not how they are connected in space; i.e., the free $R$-module generated by a set of $N$ apples is isomorphic to the free $R$-module generated by a set of $N$ oranges, or anything else. It is in the definition of the actual incidence numbers that one must include the specific details of how the geometrical object is constructed from its vertices, edges, faces, etc. That is, the topological content of the geometrical complex is contained solely in the definition of its boundary operator.

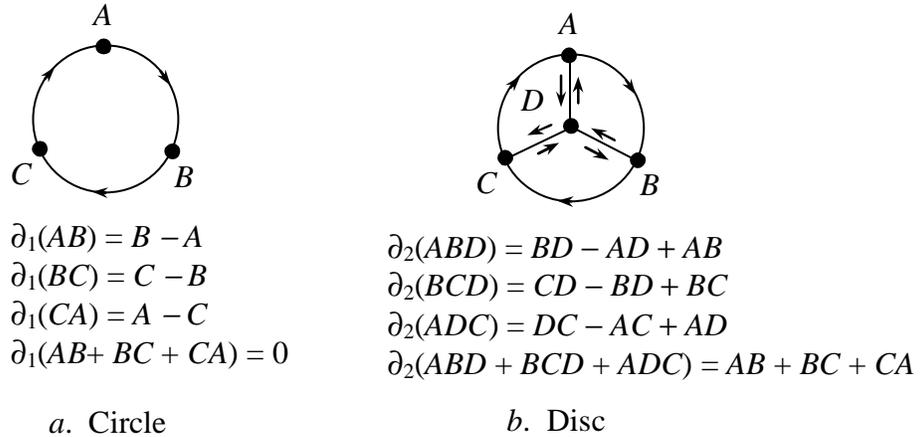

$\partial_1(AB) = B - A$

$\partial_1(BC) = C - B$

$\partial_1(CA) = A - C$

$\partial_1(AB + BC + CA) = 0$

*a.* Circle

$\partial_2(ABD) = BD - AD + AB$

$\partial_2(BCD) = CD - BD + BC$

$\partial_2(ADC) = DC - AC + AD$

$\partial_2(ABD + BCD + ADC) = AB + BC + CA$

*b.* Disc

Figure 1. Some elementary triangulations.

*e. Examples.* – In order to completely specify a geometric complex, one must define both the basic chains in each dimension and the boundary operator that relates them. The process of starting with a region in space and expressing it as the geometric realization of a simplicial complex is referred to as *triangulation*, at least when the basic simplexes are triangular (tetrahedral, etc.). For instance, we illustrate this process in the elementary cases of simplicial complexes that represent a circle and a disc in Figure 1.

If we let $\sigma_0(i)$, $i = 1, 2, 3$ be $A$, $B$, and $C$, resp., while $\sigma_1(a)$, $a = 1, 2, 3$ represent $AB$, $AC$, and $BC$, resp. then the boundary operators $\partial_1$ and $\partial_0$ for the circle (Fig. 1.*a*) will have the incidence matrices:

$$\partial_a^i = \begin{bmatrix} -1 & 1 & 0 \\ -1 & 0 & 1 \\ 0 & -1 & 1 \end{bmatrix}, \qquad [\partial_0]_i = [0, 0, 0] \qquad (4.21)$$

Hence, we can express the circle in terms of the set of generating simplexes $\{A, B, C, AB, BC, CA\}$ when it is given the boundary operators that are defined in (4.21). We can also say that is consists of the three 0-simplexes $\sigma_0(i)$, $i = 1, 2, 3$, along with the 1-chain:

$$c_1 = \sigma_1(1) + \sigma_1(2) + \sigma_1(3), \qquad (4.22)$$

and the boundary operator that couples them.

There is clearly just the one elementary 1-cycle that is generated by the 1-simplexes, namely:

$$z_1 = c_1, \qquad (4.23)$$

Since the only 1-boundary is 0, it will also be the single generator of $H_1(\mathcal{S}_1; R) \cong R$.

All 0-chains will be 0-cycles for the $\partial$ operator; in particular, all of the generating 0-simplexes are 0-cycles. The three elementary boundary 0-chains $b_0(1) \equiv \sigma_0(2) - \sigma_0(1)$, $b_0(2) \equiv \sigma_0(3) - \sigma_0(2)$, $b_0(3) \equiv \sigma_0(1) - \sigma_0(3)$ are not independent, since one has $b_0(1) + b_0(2) + b_0(3) = 0$. Hence, $B_0(\mathcal{S}_1; R)$ will have two generators, while $B_0(\mathcal{S}_1; R)$ will have three. $H_0(\mathcal{S}_n; R)$ will then have one generator: $H_0(\mathcal{S}_n; R) \cong R$, which is consistent with the fact that the circle is path-connected.

If we let $\sigma_0(i)$, $i = 1, 2, 3, 4$ be $A$, $B$, $C$, and $D$, resp., while $\sigma_1(a)$, $a = 1, \ldots, 6$ represent $AB$, $AC$, $AD$, $BC$, $BD$, $CD$, resp., and $\sigma_2(b)$, $b = 1, 2, 3$, represent $ABD$, $BCD$, $ADC$, resp. then the boundary operators $\partial_1$ and $\partial_0$ for the disc (Fig. 1.*b*) will have the incidence matrices:

$$\partial_b^a = \begin{bmatrix} 1 & 0 & 0 \\ 0 & 0 & -1 \\ -1 & 0 & 1 \\ 0 & 1 & 0 \\ 1 & -1 & 0 \\ 0 & 1 & -1 \end{bmatrix}, \qquad \partial_a^i = \begin{bmatrix} -1 & -1 & -1 & 0 & 0 & 0 \\ 1 & 0 & 0 & -1 & -1 & 0 \\ 0 & 1 & 0 & 1 & 0 & -1 \\ 0 & 0 & 1 & 0 & 1 & 1 \end{bmatrix},$$

(4.24)

$$[\partial_0]_i = [0, 0, 0, 0] \ .$$

Hence, we can express the disc as a 2-dimensional chain complex that is generated by the simplexes $\{A, B, C, D, AB, AC, AD, BC, BD, CD, ABD, BCD, ADC\}$ when it is given the boundary operators that are defined in (4.24). Similarly, one triangulates it into simply the 2-chain:

$$c_2 = \sigma_2\,(1) + \sigma_2\,(2) + \sigma_2\,(3), \tag{4.25}$$

whose boundary is:

$$\partial c_2 = b_1\,(0) \equiv \sigma_1\,(1) + \sigma_1\,(2) + \sigma_1\,(3), \tag{4.26}$$

since the contributions from the internal links cancel in pairs. In fact, $b_1\,(0)$ is the sum of three generating boundaries:

$$b_1\,(0) = b_1\,(1) + b_1\,(2) + b_1\,(3), \tag{4.27}$$

which are defined by $b_1\,(1) = BD + DA + AB$, $b_1\,(2) = CD + DB + BC$, $b_1\,(3) = DC + CA + AD$ . Since all of the 1-cycles are 1-boundaries, that will make $H_1\,(\mathcal{S}_n\,;R) = 0$. (This relates to the fact that the disc is simply connected.)

Since there are no 3-chains, $H_2\,(\mathcal{S}_n\,;R) \cong Z_2\,(\mathcal{S}_n\,;R)$, which has no generators, either. The fact that there are no 2-cycles in the disc comes from the fact that it has a boundary. Interestingly, one can make that boundary go away by "doubling" the disc – i.e., taking two disjoint copies of the disc, giving the second the opposite orientation, and identifying the pairs of corresponding boundary points to single points of the "quotient space." The doubling of the complex that triangulates a disc will then be straightforward. The resulting complex will triangulate a 2-sphere, and since a sphere has no boundary, the sum of the two elementary 2-chains for the discs (i.e., hemispheres) will be a non-vanishing 2-cycle that represents the single generator for the two-dimensional homology module.

Since the disc is path-connected, $H_0\,(\mathcal{S}_n\,;R)$ will have one generator and will therefore be isomorphic to $R$.

One can represent all of the compact surfaces without boundary by means of identifications of the edges of a square, which then leads to triangulations (more precisely, *quadrangulations*) of them that allow one to also express their homology modules (see Massey [**15**]).

*f. Cohomology* –  Cohomology works the same for geometric realizations of abstract complexes as it did for the abstract complexes.  That is, the $R$-module $C^k (\Lambda_m ; R)$ of $k$-cochains with values in $R$ will again be dual to $C_k (\Lambda_m ; R)$ and will thus represent linear functionals on $k$-chains that take their values in $R$.  The evaluation of a $k$-cochain $c^k$ on a $k$-chain $c_k$ will again be represented by $<c^k , c_k>$.  The coboundary operator $\delta : C^k (\Lambda_m ; R) \rightarrow C^{k+1} (\Lambda_m ; R)$ will still be the transpose of the boundary operator $\partial$ :

$$<\delta c^k , c_{k+1}> = <c^k , \partial c_{k-1}>. \tag{4.28}$$

Its kernel will be denoted by $Z^k (\Lambda_m ; R)$, and its elements will be referred to as *k-cocycles.*  Its image will be denoted by $B^{k+1} (\Lambda_m ; R)$, and its elements will be referred to a $(k+1)$-coboundaries.  The quotient $R$-modules $H^k (\Lambda_m ; R) = Z^k (\Lambda_m ; R) / B^k (\Lambda_m ; R)$ will be referred to as the $k$-dimensional cohomology modules, and their elements will be equivalence classes of $k$-cocyles that differ by a $k$-coboundary; i.e., they are cohomologous.

## 5.  Statics.
– Statics is a topic that really becomes topologically interesting only for systems of forces that act upon systems of points and displacements, since the basic equations of statics (viz., the vanishing of the resultant forces and force moments about some point) do not take on their homological context until one has more than one point to consider.  Indeed, even a system of $N$ non-interacting points will still define only a 0-dimensional complex, and it is only the introduction of an "interaction model" that will extend the dimensions of the complex to one.

In this section, we define a (static) *mechanical network* ([1]) to consist of a *structural network*, which is the geometric realization of a one-dimensional simplicial complex $\mathcal{S}_1$ in an affine space $A^n$, along with a "position-displacement complex" that consists of a position 0-cochain and a displacement 1-cochain, both of which have vector coefficients and are connected to each other by a coboundary, and a "force complex" that consists of the forces that act on the nodes and branches of the network, which are then represented by a 0-chain and a 1-chain with vector coefficients, and a coboundary operator that connects them.

We will then show that the equilibrium equations for force will then take on the character of saying that the 0-chain of (resultant) external forces that act upon the nodes must equal the boundary of the 1-chain of internal forces that act within the links.  That will then become an analogue of Kirchhoff's law of currents.  The equilibrium of force moments about some point is slightly more problematic, although it still relates to homology.

We shall find that a mechanical network is one step up from an electrical network in terms of complexity, in the sense that the "signals" that one associates with the nodes and branches of the abstract network will be vector-valued functions of time, instead of scalar-valued ones.  However, the set of such vector-valued functions has a natural structure as an Abelian group, but not a ring, since the multiplication of vectors is undefined in general. Hence, some care must be taken to define the relevant modules and operators.

The first thing that we shall find is that the relationship between the absolute positions of the points of the network in space and their relative displacements is analogous to the relationship between the absolute voltages at the nodes of an electrical network and the voltage drops across the branches. We will then find that the force distribution that acts on a network of points behaves like the currents and charges in an electrical network. In particular, both of Kirchhoff's laws of electrical circuits have their analogues in terms of mechanical networks.

*a. The position-displacement complex of a mechanical network.* – Suppose that $\mathcal{L}_1 = \{L_0, L_1, \partial\}$ is a geometric realization of a one-dimensional abstract simplicial complex $\mathcal{S}_1 = \{S_0, S_1, \partial\}$ in an $n$-dimensional affine space $A^n$.

If one associates each point $x(i)$ in $L_0$ with a *position vector* $\mathbf{x}(i)$ in $\mathbb{R}^n$ by way of the map (4.9) (relative to some point $O$ in $A^n$ that will play the role of origin) then one can think of that association as a $\mathbb{Z}$-linear map $\mathbf{x} : C_0(\mathcal{L}_1) \to \mathbb{R}^n$, $i \mapsto \mathbf{x}(i)$ from the $\mathbb{Z}$-module of 0-chains to vectors in $\mathbb{R}^n$, which one regards as a $\mathbb{Z}$-module by the inclusion of $\mathbb{Z}$ in the scalar field $\mathbb{R}$. Hence, if $c_0 = \sum_{i=1}^{r_0} m(i)\,\sigma_0(i)$ is a 0-chain with integer coefficients then:

$$\mathbf{x}(c_0) = \sum_{i=1}^{r_0} m(i)\,\mathbf{x}[\sigma_0(i)] = \sum_{i=1}^{r_0} m(i)\,\mathbf{x}(i)\,. \tag{5.1}$$

We can then think of the *position distribution* of the points of $L_0$ as being a 0-cochain with values in $\mathbb{R}^n$: i.e., $\mathbf{x} \in C^0(\mathcal{L}_1 ; \mathbb{R}^n)$. Hence, if $\{\sigma^0(i), i = 1, \dots, r_0\}$ is the set of generators of $C^0(\mathcal{L}_1 ; \mathbb{Z})$ that is reciprocal to the set of generators $\sigma_0(i)$ for $C_0(\mathcal{L}_1 ; \mathbb{Z})$ – so $<\sigma^0(i), \sigma_0(j)> = \delta_j^i$ – then we can express $\mathbf{x}$ in the form:

$$\mathbf{x} = \sum_{i=1}^{r_0} \mathbf{x}(i)\,\sigma^0(i)\,. \tag{5.2}$$

Actually, we are being slightly imprecise in our reference to $\mathbf{x}$ as a 0-cochain with values in $\mathbb{R}^n$, since $\mathbb{R}^n$ is not actually a ring. More precisely, we are regarding $\mathbb{R}^n$ as a $\mathbb{Z}$-module and defining $C^1(\mathcal{L}_1 ; \mathbb{R}^n)$ to be the tensor product $\mathbb{R}^n \otimes C^1(\mathcal{L}_1 ; \mathbb{Z})$, which is a $\mathbb{Z}$-module whose elements consist of $\mathbb{Z}$-linear functionals on the tensor product $\mathbb{R}^n \otimes C_1(\mathcal{L}_1 ; \mathbb{Z})$. In that light, we should really write (5.6) in the form:

$$\mathbf{x} = \sum_{i=1}^{r_0} \mathbf{x}(i) \otimes \sigma^0(i), \tag{5.3}$$

but hopefully the omission of the tensor product symbol will not make things too confusing.

Since we know from (4.6) that the 0-chain $\sigma_0(j) - \sigma_0(i)$, which is also $x(j) - x(i)$, can be associated with a displacement vector $\mathbf{s}(i, j)$, we can also think of a 1-simplex $\sigma_1(a)$ that is associated with $(i, j)$ as being associated with a displacement vector in $\mathbb{R}^n$:

$$\mathbf{s}(i, j) = \mathbf{s}[\sigma_1(i, j)]. \tag{5.4}$$

Hence, if one extends by linearity to an arbitrary 1-chain $c_1 = \sum_{a=1}^{r_1} m(a)\,\sigma_1(a)$:

$$\mathbf{s}(c_1) = \sum_{a=1}^{r_1} m(a)\mathbf{s}[\sigma_1(a)] = \sum_{a=1}^{r_1} m(a)\mathbf{s}(a) \tag{5.5}$$

then the displacement distribution of $S$ will define a $\mathbb{Z}$-linear map $\mathbf{s}: C_1(\mathcal{L}_1; \mathbb{Z}) \to \mathbb{R}^n$, $c_1 \mapsto \mathbf{s}(c_1)$ ; i.e., a 1-cochain with values in $\mathbb{R}^n$, so $\mathbf{s} \in C^1(\mathcal{L}_1; \mathbb{R}^n)$. One can then write:

$$\mathbf{s} = \sum_{a=1}^{r_1} \mathbf{s}(a)\,\sigma^1(a). \tag{5.6}$$

Unlike the association of position vectors, which requires an otherwise-arbitrary choice of origin for the affine space, the displacement 1-chain requires no such choice, and is therefore unambiguous in that sense.

This time, the bilinear pairing is between $C^1(\mathcal{L}_1, \mathbb{R}^n)$ and $C_1(\mathcal{L}_1, \mathbb{Z})$, not $C^1(\mathcal{L}_1, \mathbb{R}^n)$ and $C_1(\mathcal{L}_1, \mathbb{R}^n)$. In order to deal with the latter case, we would also need a bilinear functional on $\mathbb{R}^n \times \mathbb{R}^n$. In a later section on energetics, we shall address that issue, but for now, we confine ourselves to the pairing in question. The coboundary operator $\Delta$: $C^0(\mathcal{L}_1, \mathbb{R}^n) \to C^1(\mathcal{L}_1, \mathbb{R}^n)$ is then defined by its action on 1-chains (with coefficients in $\mathbb{Z}$) [1]:

$$<\Delta\mathbf{c}^0, c_1> \equiv <\mathbf{c}^0, \partial c_1>. \tag{5.7}$$

In particular, with $\mathbf{x}$ as in (5.2), one will have:

---

[1]  Since we will eventually need the $\delta$ symbol to represent variations, from now on, we shall use the symbol $\Delta$ to represent the coboundary operator.

$$\Delta \mathbf{x} = \sum_{i=1}^{r_0} \mathbf{x}(i)\,\Delta\sigma^0(i) \;=\; \sum_{a=1}^{r_1}\left[\sum_{i=1}^{r_0}\mathbf{x}(i)\,\partial_a^i\right]\sigma^1(a)\,. \tag{5.8}$$

For each $a$, the expression in square brackets will take the form:

$$\Delta\mathbf{x}(a) = \sum_{i=1}^{r_0}\mathbf{x}(i)\,\partial_a^i \;=\; \mathbf{x}(j_a) - \mathbf{x}(i_a)\,, \tag{5.9}$$

if $j_a$ is the index of the final node of $\sigma_1(a)$ and $i_a$ is its initial node.

The displacement $\mathbf{s}(a)$ that is associated with a basic simplex $\sigma_1(a)$ whose boundary is $\sigma_0(j_a) - \sigma_0(i_a)$ will then be the difference of the vectors $\mathbf{x}(i_a)$, $\mathbf{x}(j_a)$ that are associated with the 0-simplexes $\sigma_0(i_a)$, $\sigma_0(j_a)$ , so:

$$< \mathbf{s},\,\sigma_1(a)> = \mathbf{x}(j_a) - \mathbf{x}(i_a) \;=\; <\mathbf{x},\,\partial\sigma_1(a)> = \;<\Delta\mathbf{x},\,\sigma_1(a)>.$$

Since that is true for all $a$, that will imply that:

$$\mathbf{s} = \Delta\mathbf{x}\,. \tag{5.10}$$

This also says that $\mathbf{s}$ is a 1-coboundary, and therefore a 1-cocycle:

$$\Delta\mathbf{s} = 0. \tag{5.11}$$

If we refer to $\mathbf{s}$ as the *displacement 1-coboundary* then, from the basic property of coboundaries, we will have:

**Theorem:**

*The displacement 1-cochain of a network of points will vanish whenever it is evaluated on a 1-cycle. In particular, it will vanish around any loop.*

This is the mechanical analogue of Kirchhoff's law of voltages then, so one can see that the arbitrariness in the choice of origin $O$ in $A^n$ corresponds to the arbitrariness in the choice of 0 for voltages. Similarly, it is only the relative positions (i.e., displacements) that are well-defined, in the same way that only voltage differences are well-defined.

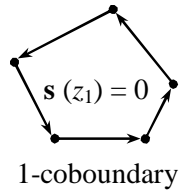

1-coboundary

Figure 2. – Relationship between $\mathbf{s}$ being a 1-coboundary
and the polygon rule for vector addition.

Basically, the fact that **s** is a 1-coboundary means that one is dealing with the "polygon rule" for vector addition, which we illustrate in Fig. 2 above.

We also find that when the displacement 1-coboundary $\Delta\mathbf{x}$ is evaluated on a path $l = \sum_{i=1}^{p} \sigma_1(i)$, so $\partial\, l = \sigma_0\,(p+1) - \sigma_0\,(1)$, we will get:

$$< \Delta\mathbf{x}, l > = < \mathbf{x}, \partial\, l > = \mathbf{x}\,(p+1) - \mathbf{x}\,(1) = \mathbf{s}\,(1, p+1). \qquad (5.12)$$

That is, the result is the displacement vector that takes the starting point to the end point. Clearly, that will vanish for a loop.

One can also see that:

$$< \Delta\mathbf{x}, l > = < \sum_{i=1}^{p} \Delta\mathbf{x}(i)\sigma^1(i), \sum_{j=1}^{p} \sigma_1(j) > = \sum_{i=1}^{p} \Delta\mathbf{x}(i), \qquad (5.13)$$

which is consistent with one's expectations for displacements along a path, namely, that:

$$\sum_{i=1}^{p} \Delta\mathbf{x}(i) = \mathbf{x}\,(p+1) - \mathbf{x}\,(1). \qquad (5.14)$$

We then define the *position-displacement complex* to be the quadruple $\{O, \mathbf{x}, \mathbf{s}, \Delta\}$ that represents the network of points in $\mathbb{R}^n$, along with the coboundary operator that links them.

The effect of changing $O$ to $O' = O + \mathbf{a}$ will be to change **x** to $\mathbf{x}' = \mathbf{x} - \mathbf{a}$, while **s** will remain unaffected, since $\mathbf{x}\,(i) - \mathbf{x}\,(j)$ will always go to:

$$[\mathbf{x}\,(i) - \mathbf{a}] - [\mathbf{x}\,(j) - \mathbf{a}] = \mathbf{x}\,(i) - \mathbf{x}\,(j).$$

If we define:

$$\mathbf{a}^0 = \sum_{i=1}^{r_0} \mathbf{a}\,\sigma^0(i) \qquad (5.15)$$

then since $\mathbf{s}' = \Delta\mathbf{x} = \Delta\mathbf{x} - \Delta\mathbf{a}^0 = \mathbf{s}$, we must conclude that:

$$\Delta\mathbf{a}^0 = 0. \qquad (5.16)$$

One can verify this directly by forming:

$$\Delta\mathbf{a}^0 = \sum_{i=1}^{r_0} \mathbf{a}\,\Delta\sigma^0(i) \qquad (5.17)$$

and evaluating it on any basic 1-simplex $\sigma_1\,(a)$:

$$<\Delta\mathbf{a}^0, \sigma_1\,(a)> = <\mathbf{a}^0, \partial\sigma_1\,(a)> = \sum_{i=1}^{r_0}[\mathbf{a}<\sigma^0(i), \sum_{j=1}^{r_0}\partial_j^a\,\sigma_0(j)>] = \mathbf{a}\sum_{i=1}^{r_0}\sum_{j=1}^{r_0}\partial_j^a\,\delta_i^j$$
$$= \mathbf{a}\,(1-1) = 0.$$

We can then say that:

$$\mathbf{x}' - \mathbf{x} = (0\text{-cocycle}) . \tag{5.18}$$

This situation is closely analogous to the one that one encounters in gauge field theories, in which one has, for example, an electromagnetic potential 1-form $A$ whose exterior derivative $F = d_\wedge A$ represents the electromagnetic field strength 2-form. Changing $A$ by adding any closed 1-form $A \to A + \psi\,(d_\wedge\psi = 0)$ will not alter $F$, and one calls $A$ a choice of *gauge* for $F$, while the transformation $A \to A + \psi$ is referred to as a *gauge transformation* (of the second kind). Hence, one sees that a choice of absolute position (i.e., a choice of $O$) can be regarded as a kinematic analogue of a choice of choosing a gauge for an electromagnetic field. We shall return to discuss the latter in Part II of this series of articles, which will be concerned with continuous systems of points.

We can extend a position-displacement complex to its cone relative to a reference point in space $O$. The absolute position vector $\mathbf{x}\,(O)$ that is associated with $O$ is zero, by definition, and the links include all of the pairs $(0, i)$ for $i \neq 0$, where $0$ is the index for $O$. That will have the effect of making:

$$\mathbf{x}\,(O) = 0, \qquad \mathbf{s}\,(0, i) \equiv \mathbf{r}\,(i) \equiv \mathbf{x}\,(i), \qquad \text{for all } i \neq O. \tag{5.19}$$

In this definition, we have included the definition of the *position vector field* $\mathbf{r}\,(i)$ (also called the *radius vector field*) of the nodes $i$ with respect to the point $O$. Hence, each $\mathbf{r}\,(i)$ will be the coefficient of $\sigma^1\,(O, i)$ in the extended complex $\{O, \overline{\mathbf{x}}, \overline{\mathbf{s}}, \overline{\Delta}\}$ with:

$$\overline{\mathbf{x}} = \mathbf{x} + \mathbf{x}\,(O) = \mathbf{x}, \qquad \overline{\mathbf{s}} = \mathbf{s} + \sum_{i=1}^{r_0}\mathbf{r}(i)\,\sigma^1(0; i), \qquad \overline{\Delta} = \Delta. \tag{5.20}$$

We shall such a complex a *position-displacement complex with a distinguished point.* We illustrate a typical situation in Fig. 3. For such a complex, the absolute positions of each node are defined to be positions relative to $O$; i.e., the displacement that takes $O$ to that node.

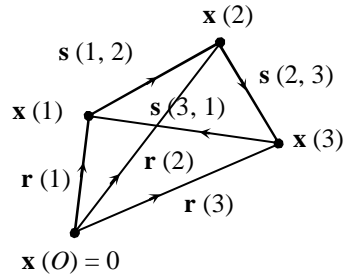

Figure 3. A position-displacement complex with a distinguished point $O$.

One can see that when one shifts the reference point $O$ to $O' = O + \mathbf{a}$, so $\mathbf{x}$ will go to $\mathbf{x}' = \mathbf{x} - \mathbf{a}$, the only displacements that get associated with links that will change as a result of that shift will be the ones that are associated with links that are incident on $O$ :

$$\mathbf{r}'(i) = \mathbf{r}(i) - \mathbf{a} . \tag{5.21}$$

Otherwise:

$$\mathbf{s}'(a) = \mathbf{s}(a) \quad \text{for all } a . \tag{5.22}$$

*b. The force complex of a mechanical network.* – There are two types of forces that can act upon a network of points in space: *(resultant) external forces* $\mathbf{F}(i)$, which act upon the nodes, and *internal forces* $\mathbf{F}(a)$, which act inside the links. If we regard forces as covectors, which then belong to $\mathbb{R}^{n*}$, then the external forces will define a 0-chain with coefficients in $\mathbb{R}^{n*}$ by way of:

$$\mathbf{F}_{\text{ext}} = \sum_{i=1}^{r_0} \mathbf{F}(i) \, \sigma_0(i) , \tag{5.23}$$

so $\mathbf{F}_{\text{ext}} \in C_0(\mathcal{L}_1 ; \mathbb{R}^{n*})$.

If one takes the augmented boundary of $\mathbf{F}_{\text{ext}}$ then one will get:

$$\partial^{\#} \mathbf{F}_{\text{ext}} = \sum_{i=1}^{r_0} \mathbf{F}(i) . \tag{5.24}$$

Hence, it represents the resultant of all the external forces that act at each node.

Similarly, the internal forces define a 1-chain with values in $\mathbb{R}^{n*}$:

$$\mathbf{F}_{\text{int}} = \sum_{a=1}^{r_1} \mathbf{F}(a) \, \sigma_1(a) , \tag{5.25}$$

so $\mathbf{F}_{\text{int}} \in C_1(\mathcal{L}_1 ; \mathbb{R}^{n*})$.

The boundary of $\mathbf{F}_{\text{int}}$ is then:

$$\partial \mathbf{F}_{\text{int}} = \sum_{a=1}^{r_1} \mathbf{F}(a) \partial \sigma_1(a) = \sum_{i=1}^{r_0} \left( \sum_{a=1}^{r_1} \mathbf{F}(a) \partial_i^a \right) \sigma_0(i) , \tag{5.26}$$

so the coefficient of the $i^{\text{th}}$ node $\sigma_0(i)$ will be:

$$\partial \mathbf{F}_{\text{int}}(i) = \sum_{a=1}^{r_1} \mathbf{F}(a) \partial_i^a , \tag{5.27}$$

which is the sum of all internal forces in the links that are incident on $\sigma_0(i)$. Hence, this is analogous to the sum of the currents at a node in an electrical circuit. One can then define the resultant internal force distribution $\mathbf{F}_{\text{res},\,i}$ to be the 0-chain with coefficients in $\mathbb{R}^{n*}$ :

$$\mathbf{F}_{\text{res},\,i} = \partial\, \mathbf{F}_{\text{int}}\,. \tag{5.28}$$

In symbols: $\mathbf{F}_{\text{res},i} \in B_0\,(\mathcal{L}_1\,;\,\mathbb{R}^{n*})$ .

Furthermore, since the internal forces are assumed to act longitudinally in the links, each internal force will be collinear with the displacement $\mathbf{s}\,(a)$ that is associated with that link. Hence, the internal forces are also defined by scalar quantities $f\,(a)$ that make:

$$\mathbf{F}\,(a) = f\,(a)\; \hat{\mathbf{s}}\,(a)\,. \tag{5.29}$$

(The displacement vector $\mathbf{s}\,(a)$ has been converted to a covector and normalized to have unit length in this.) When the sign of $f\,(a)$ is positive, one says that the link $\sigma_1(a)$ is in a state of *tension*, because the force will tend to elongate the link, and when the sign of $f\,(a)$ is negative, a state of *compression* will exist in the link, because the tendency of the force is to reduce the length of the link.

We then define the *force complex* of the mechanical network to be the triple $\{\mathbf{F}_{\text{ext}}, \mathbf{F}_{\text{int}}, \partial\}$, which consists of the 0-chain $\mathbf{F}_{\text{ext}}$ and the 1-chain $\mathbf{F}_{\text{int}}$ , which both have coefficients in $\mathbb{R}^{n*}$, along with the boundary operator that couples them. One can also define the *extended force complex* to be the quadruple $\{\mathbf{F}_{\text{ext}}\,(\infty)\; \mathbf{F}_{\text{int}}\,(\infty),\, \infty,\, \partial\}$, in which $\infty$ is a "point at infinity," with:

$$\mathbf{F}_{\text{ext}}\,(\infty) = \mathbf{F}_{\text{ext}} + \left[\sum_{i=1}^{r_0}\mathbf{F}_{\text{ext}}\,(i)\right]\sigma_0\,(\infty)\,, \qquad\qquad \mathbf{F}_{\text{int}}\,(\infty) = \mathbf{F}_{\text{int}} + \sum_{i=1}^{r_0}\mathbf{F}_{\text{ext}}\,(i)\,\sigma_1\,(i,\infty)\,, \tag{5.30}$$

and $\partial$ is extended by $\partial\sigma_1(i,\infty) = \sigma_0(\infty) - \sigma_1(i)$ for every $i$. That will then make:

$$\partial\mathbf{F}_{\text{int}}\,(\infty) = \partial\mathbf{F}_{\text{int}} + \sum_{i=1}^{r_0}\mathbf{F}_{\text{ext}}\,(i)\,\partial\sigma_1(i,\infty) = \left[\sum_{i=1}^{r_0}\mathbf{F}_{\text{ext}}\,(i)\right]\sigma_0\,(\infty)\,. \tag{5.31}$$

Hence, the vanishing of $\partial\mathbf{F}_{\text{int}}\,(\infty)$ is equivalent to the vanishing of the resultant of the external forces.

Notice that since the directions of the external forces $\mathbf{F}_{\text{ext}}\,(i)$ do not have to be parallel, one is not dealing with the extension of the affine space $A^n$ to the projective space $\mathbb{R}\mathrm{P}^n$, which would require that lines that connect to the same point at infinity must be parallel. Rather, one is dealing with the compactification of $A^n$ by a single point at infinity, which would make it an $n$-sphere.

*c. Equilibrium.* – A mechanical network is in a state of *static equilibrium* iff the resultant of the external forces vanishes, along with the sum of the internal and external forces that act at each node. Hence, the condition for static equilibrium will become:

$$\sum_{i=1}^{r_0} \mathbf{F}_{ext}(i) = 0, \qquad \mathbf{F}_{ext} = -\mathbf{F}_{res,\,i} \; ; \qquad (5.32)$$

i.e.:

$$\partial^{\#}\mathbf{F}_{ext} = 0, \qquad \mathbf{F}_{ext} = -\partial\,\mathbf{F}_{int} \,. \qquad (5.33)$$

When expressed in terms of the individual nodes, the second of equations (5.33) will become:

$$\mathbf{F}_{ext}(i) = -\sum_{a=1}^{r_1} \mathbf{F}(a)\,\partial_i^a \,. \qquad (5.34)$$

If one interprets the external forces $\mathbf{F}_{ext}(i)$ that act upon the nodes as the analogues of the electrical charges that accumulate in an electrical circuit then equations (5.33) will become the mechanical analogue of Kirchhoff's law of currents, when extended by $\partial^{\#}$:

**Theorem (force equilibrium):**

*If a force distribution on a network of points is in static equilibrium then $\mathbf{F}_{ext}$ will be a 1-boundary and $\mathbf{F}_{int}(\infty)$ will be a 1-cycle.*

The fact that the necessary conditions are not sufficient is due to the fact that one must also require the vanishing of the total force moment about some point, which we will discuss shortly.

When a structure is rigid, the internal forces will not contribute to its motion, and only the vanishing of the resultant of the external forces will be necessary for static equilibrium. In that case, one can require that either:

$$\partial^{\#}\mathbf{F}_{ext} = 0 \qquad \text{or} \qquad \partial\mathbf{F}_{int}(\infty) = 0 \,. \qquad (5.35)$$

*d. Internal stress states.* – Notice that it is entirely possible for equilibrium to exist in a mechanical network when there are no external forces acting upon it, but there are still internal forces acting in the branches. In that case, from (5.33), one would be looking for non-zero solutions of the equilibrium condition:

$$\partial\,\mathbf{F}_{int} = 0 \,. \qquad (5.36)$$

In other words, the internal force distribution would take the form of a 1-cycle.

Since there are no 2-chains in our complex, there will be no (non-zero) 1-boundaries, and any 1-cycle will define a generator of the one-dimensional homology module $H_1(\mathcal{L}_1 ; \mathbb{R}^{n^*})$. Hence, the existence of internal stresses can be due to purely-topological sources when $H_1(\mathcal{L}_1 ; \mathbb{R}^{n^*})$ is not merely the 0-module. In practice, though, there are other ways that internal stresses can show up, such as thermal stresses due to expansion and contraction, as well as manufacturing stresses.

An example of a network of non-zero internal stresses that exist in the absence of external loads is given in Fig. 6 below, which depicts a triangle with rigid sides that has tensed elastic members attached to its vertices and a central vertex, along with the force diagrams for each vertex.

An important property of the network that is defined in Fig. 6 is that it represents the projection of the edge complex (1-skeleton) of a tetrahedron onto its base. In fact, James Clerk Maxwell showed [**17**] that if a network is the projection of the 1-skeleton of a convex polyhedron then it will admit non-zero internal stress states. (See also [**18**].)

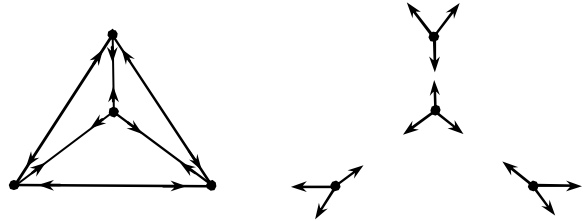

Figure 6. – A mechanical network in which internal
stresses can exist in the absence of external loads.

*e. Statically-determinate and statically-indeterminate frameworks.* – A typical problem of statics is to compute the internal stresses in a framework for a given distribution of external loads. The equations of static equilibrium (5.33) can then be regarded as a system of linear equations for the $r_1$ unknowns $\mathbf{F}(a)$. Since there are $r_0$ nodes, and therefore $r_0$ external forces $\mathbf{F}(i)$, the questions immediately arise of whether one can solve the system of equations and whether the solution will be unique.

When $r_0 = r_1$, one says that the framework is *statically determinate* because the equations can be solved for a unique distribution of internal forces (assuming that those equations are also consistent). However, it is more typical in engineering practice to have more links than nodes, so $r_0 < r_1$, and one says that such a structure is *statically indeterminate.* What makes that interesting from the standpoint of homology is that since the linear map in question is the boundary map for a one-dimensional complex, the question of whether it is a linear injection comes down to the question of whether its kernel vanishes. Since the kernel of $\partial$ is $Z_1(\mathcal{L}_1 ; \mathbb{R}^{n^*})$, which is then $H_1(\mathcal{L}_1 ; \mathbb{R}^{n^*})$, since there are no 1-boundaries but 0, one sees that statically-indeterminate structures will typically admit non-zero internal stress states. Hence, the issues of static determinacy and the existence of internal stress states are both related to the same question in homology, namely, the existence of non-trivial generators to the module $H_1(\mathcal{L}_1 ; \mathbb{R}^{n^*})$.

As an example of the difference, we use the triangle and the projected tetrahedron above with loads applied to the external vertices, which we illustrate in Fig. 7:

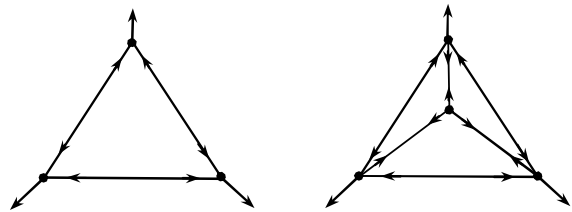

Figure 7. – Statically-determinate and statically-indeterminate frameworks.

Although static indeterminacy sounds like a bad thing, it tends to be the rule, more than the exception, in structural engineering practice, since the redundant structural members, which might carry vanishing internal forces in equilibrium, can nonetheless serve to stabilize a structure under the perturbations to the unloaded equilibrium state that one must expect from the application of varying loads.

*f. The force moment complex of a mechanical network.* – The concept of the moment of a tangent vector at one point with respect to a chosen reference point can be treated in terms of either vector cross products on $\mathbb{R}^3$ or bivectors on more general vector spaces. We will choose the latter approach because it is not as tied down to three dimensions, but we shall nonetheless point out that there is another algebraic consequence of the vector product that one must not forget, namely, that it defines the structure of a Lie algebra on $\mathbb{R}^3$. In particular, it defines the Lie algebra $\mathfrak{so}(3; \mathbb{R})$, which consists of infinitesimal generators of one-parameter subgroups of the three-dimensional Euclidian rotation group $SO(3; \mathbb{R})$, which represent elementary rotational motions.

If $V$ is an $n$-dimensional vector space then a *bivector* $\mathbf{B}$ over $V$ is a finite sum of exterior products of vectors $\{\mathbf{a}_i, \mathbf{b}_j, i, j, = 1, \ldots, p\}$ in $V$ :

$$\mathbf{B} = \sum_{i=1}^{p} \mathbf{a}_i \wedge \mathbf{b}_j . \tag{5.37}$$

We shall define the *exterior product* $\wedge$ axiomatically here, and return to it in the sequel to this article.

*i*)  If $\mathbf{a}, \mathbf{b} \in V$ then $\mathbf{a} \wedge \mathbf{b}$ belongs to a vector space $\Lambda_2 V = V \wedge V$ whose dimension is $n(n-1)/2$ .

*ii*)  The exterior product is bilinear and antisymmetric. Hence, if $\mathbf{a}, \mathbf{b}, \mathbf{c} \in V$ and $\alpha, \beta \in \mathbb{R}$ then:

$$(\alpha \mathbf{a} + \beta \mathbf{b}) \wedge \mathbf{c} = \alpha \mathbf{a} \wedge \mathbf{c} + \beta \mathbf{b} \wedge \mathbf{c}, \tag{5.38}$$

$$\mathbf{c} \wedge (\alpha \mathbf{a} + \beta \mathbf{b}) = \alpha \mathbf{c} \wedge \mathbf{a} + \beta \mathbf{c} \wedge \mathbf{b}, \tag{5.39}$$

$$\mathbf{a} \wedge \mathbf{b} = - \mathbf{b} \wedge \mathbf{a} . \tag{5.40}$$

Actually, these three conditions are not independent. For instance, (5.38), in conjunction with (5.40), will imply (5.39). Furthermore, (5.40) implies that:

$$\mathbf{a} \wedge \mathbf{a} = 0 \tag{5.41}$$

for any vector $\mathbf{a}$.

*iii*)  If $\{\mathbf{e}_i, i = 1, \ldots, n\}$ is a basis for $V$ then $\{\mathbf{e}_i \wedge \mathbf{e}_j, i, j = 1, \ldots, n\}$ will be a (redundant) basis for $\Lambda_2 V$. It is redundant because from (5.40), $\mathbf{e}_i \wedge \mathbf{e}_j$ will be proportional to $\mathbf{e}_j \wedge \mathbf{e}_i$ (by way of $-1$) and $\mathbf{e}_i \wedge \mathbf{e}_i$ will be zero in any case. Hence, the expression for a bivector $\mathbf{B}$ in terms of that basis will require a factor of $1/2$ to correct for the redundancy:

$$\mathbf{B} = \tfrac{1}{2} B^{ij} \mathbf{e}_i \wedge \mathbf{e}_j , \tag{5.42}$$

in which the sums over all $i$ and $j$ will now be regarded as implicit.

In particular, the components of $\mathbf{a} \wedge \mathbf{b}$ will be:

$$(\mathbf{a} \wedge \mathbf{b})_{ij} = a_i b_j - a_j b_i . \tag{5.43}$$

If one chooses a point $O$ in an affine space $A^n$ to serve as a reference point or origin then the map $\mathbf{s}_O : A^n \to \mathbb{R}^n$, $x \mapsto x - O$ will define the position vector field $\mathbf{r}(x) = x - O$, which will associate each point $x$ with a tangent vector $\mathbf{r}(x)$ in $T_x A^n$.

If $\mathbf{y}$ is another tangent vector to the point $x \in A^n$ that is also represented by the same symbol as the vector in $\mathbb{R}^n$ that is its image under $\mathbf{s}_x$ then the *moment of* $\mathbf{y}$ *with respect to* $O$ will be defined by:

$$\mathbf{M}_O(\mathbf{y}) = \mathbf{r}(x) \wedge \mathbf{y} , \tag{5.44}$$

which will then belong to $\Lambda_2 \mathbb{R}^n$. Hence, if $\mathbf{e}_i$ is a basis for $\mathbb{R}^n$ then the components of $\mathbf{M}_O$ $(\mathbf{y})$ with respect to that basis will be:

$$(\mathbf{M}_O(\mathbf{y}))_{ij} = x_i y_j - x_j y_i . \tag{5.45}$$

Due to the antisymmetry of the component matrix $B_{ij}$ of any bivector $\mathbf{B}$ with respect to a chosen basis, combined with the fact that infinitesimal rotations can also be represented by antisymmetric matrices, a bivector can also be regarded as the infinitesimal generator of a rotational motion. More precisely, one must raise an index of $B_{ij}$ using the scalar product $\eta$ that relates to the rotations in question and get a matrix:

$$B^i_j = \eta^{jk} B_{kj} \tag{5.46}$$

that represents the infinitesimal rotation that is associated with the bivector $\mathbf{B}$ ([1]).

In the case of "decomposable" bivectors, which take the form $\mathbf{a} \wedge \mathbf{b}$ (but not uniquely), as long as $\mathbf{a} \wedge \mathbf{b}$ is non-zero, the vectors $\mathbf{a}$ and $\mathbf{b}$ will not be collinear and will therefore span a plane in $V$. When $V$ is three-dimensional Euclidian space, the normal to the plane of $\mathbf{a}$ and $\mathbf{b}$ will be the axis of the infinitesimal rotation that is described by $\mathbf{a} \wedge \mathbf{b}$, and the orientation of that axis is usually assigned by the "right-hand screw rule," which says that the positive direction of the axis points in the direction of advance of a right-hand screw when one rotates **a** towards **b** in the plane that they span.

---

([1]) Although the association of $\mathbf{B}$ with an infinitesimal rotation can be derived in a "basis-free" way that does not appeal to component matrices, nonetheless, that association is more intuitive when it is expressed in terms of components.

For now, we shall focus upon the moment of a force covector $F$ that acts at a point $x$ with respect to a chosen reference point $O$. First, since we have defined force to be a covector, we have to define the *position covector field* $r\,(x)$ on $A^n$, its value at $x$ is obtained from the composition of the map $\mathbf{s}_O$, which takes $x$ to $\mathbf{r}\,(x)$ in $\mathbb{R}^n$, with the transpose map from $\mathbb{R}^n$ to its dual $\mathbb{R}^{n*}$, which takes the column vector $(x^1, \ldots, x^n)^{\mathrm{T}}$ to the row vector $(x_1, \ldots, x_n)$ with $x^i = x_i$ for all $i$. Since the product of the row vector with the column vector is:

$$(x_1, \ldots, x_n)(x^1, \ldots, x^n)^{\mathrm{T}} = x^i\,x_i = \delta_{ij}\,x^i\,x^j = \langle\mathbf{x}, \mathbf{x}\rangle,$$

one sees that the transpose map $\mathrm{T} : \mathbb{R}^n \to \mathbb{R}^{n*}$ coincides with the map that takes $\mathbf{x}$ to the linear functional that takes every $\mathbf{y}$ to $\langle\mathbf{x}, \mathbf{y}\rangle$, which is defined by the Euclidian scalar product.

Since the dual space $V^*$ to any vector space $V$ is still a vector space, one can form exterior products of its elements as well. If the elements of $V^*$ are referred to as covectors or *(algebraic) 1-forms on V* then the elements of $\Lambda^2 V = V^* \wedge V^*$ will be referred to as *(algebraic) 2-forms on V.*

With those preliminaries, the moment of the force covector $F$, which acts at $x$, with respect to the reference point $O$, is defined to be the 2-form:

$$M_O\,(F) = r\,(x) \wedge F. \tag{5.47}$$

This has the immediate consequence that, from the antisymmetry of the exterior product, a force that is collinear to $r\,(x)$ will produce no force moment. Indeed, if one decomposes $F$ into a sum $F_r + F_t$ of a radial component $F_r = \langle F,\ \hat{r}\,(x)\rangle\hat{r}$ and a transverse component $F_t = F - F_r$ then one will see that:

$$M_O\,(F) = r\,(x) \wedge F_t. \tag{5.48}$$

That is, only the transverse component of $F$ will produce any moment.

If $\mathbf{e}_i$ is a basis for $V$ then there will be a unique reciprocal basis $\{\theta^i, i = 1, \ldots, n\}$ for $V^*$ that is defined by:

$$\theta^i\,(\mathbf{e}_j) = \delta^i_j. \tag{5.49}$$

If the components of $r\,(x)$ and $F$ with respect to the reciprocal basis $\theta^i$ are $x_i$ and $F_i$, respectively, then the components of $M_O\,(F)$ will be:

$$M_O\,(F)_{ij} = x_i\,F_j - x_j\,F_i. \tag{5.50}$$

Like forces, there are two types of force moments to consider: external and internal. The external force moments $M\,(i)$ are applied to axes that go through the nodes of the network and define an *external force moment* 0-chain:

$$M_{\text{ext}} = \sum_{i=1}^{r_0} M(i)\,\sigma_0(i)\,. \qquad (5.51)$$

Hence, $M_{\text{ext}} \in C_0\left(\mathcal{L}_1 \,;\, \Lambda^2\,\mathbb{R}^n\right)$. Note that these external moments are not the moments of the external forces that act at the nodes about some common rotational axis, but are moments of forces that are external to the network about axes that go through the nodes. One can think of them as like the action of wrenches on nuts that are located at the nodes.

It is natural to wonder what the 2-form $\partial^{\#}M_{\text{ext}}$ might represent. Since it basically equals the sum of all the external moments that act at each node of the network, one finds that unlike the total external force that acts upon a mechanical network, the total external force moment is not as well-defined, since the external moments might be about different points. Typically, one focuses on just the resultant external force moment that is applied to a given node.

In order to define the 1-chain of force moments about a common reference point $O$ (really a common axis through it), one must first extend the original network to a network with a distinguished point $O$, as was described above. However, it is also common for the distinguished point to be one of the original nodes $i$, so the only potential extension in that case might involve adding the position vectors $\{\mathbf{r}\,(j),\ j \neq i\}$, if necessary, and extending $\partial$ accordingly.

One then defines the *force moment about O* to be the 1-chain:

$$M_O = \sum_{i=1}^{r_0} [r(i) \wedge F(i)]\,\sigma_1(O,i) \equiv \sum_{i=1}^{r_0} M_O(i)\,\sigma_1(O,i)\,. \qquad (5.52)$$

Hence, its boundary will be:

$$\partial M_O = \sum_{i=0}^{r_0} \left[\sum_{j=0}^{r_0} M_O(j)\,\partial_j^i\right]\sigma_0(i)\,,$$

in which $i = 0$ corresponds to the point $O$. However, since:

$$M_O(0) = x(0,0) \wedge F(0) = 0,$$

the coefficient of $\partial M_O$ at $i = 0$ will be:

$$\partial M_O(0) = \sum_{j=1}^{r_0} M_O(j)\,\partial_j^i\ , \qquad (5.53)$$

which is the resultant of the moments with respect to $i = 0$ of the external forces that act at all the other nodes besides $i = 0$.

One can then summarize the equilibrium condition for force moments about $i$ in the form:

$$M_{\text{ext}}(0) = -\partial M_O(0)\,. \qquad (5.54)$$

That is: the sum of the internal and external moments that act at $O$ must vanish.

This condition differs from the corresponding condition for force equilibrium by the fact that it is only true for the node $O$.

**6. Kinematics.** – Kinematics is concerned with the description of motion in space, so it is basically a problem in geometry for the motion of points, but when one is concerned with the motion of extended matter, one must consider that spatially-extended regions in which one finds matter can not only move through space collectively, but also deform in the process. One might consider the difference between playing "catch" with a baseball and playing that game with a water balloon. Typically, one needs to factor the total motion through space into a collective rigid motion, such as the motion of a "body frame" that is placed at the center of mass, and a pure deformation about that center of mass. However, the decomposition of the total motion in that way is not by any means unique. Hence, what we will be calling a "deformation" of a structural complex will amount to a complex that represents the total motion of that initial structural complex.

Eventually, we will find that successive states of deformation of a structural complex can be connected by simplexes of varying dimensions that will define what we will call a "kinematical complex," which consists of all the simplexes that connected vertices and links of successive states of deformation of the structural complex in time.

*a. Finite deformations of structural complexes.* – Suppose that one has a network of points $\mathcal{L}_1$ (0) in an affine space $A^n$ that is associated with a position-displacement complex $N$ (0) = $\{O, \mathbf{x}_0, \mathbf{s}_0, \Delta_0\}$ that consists of a position 0-cochain $\mathbf{x}_0$ that describes the locations of the nodes $\sigma_0$ $(i)(0)$, $i = 1, \ldots, r_0$ with respect to a reference point $O$, a displacement 1-cochain $\mathbf{s}_0$ that describes the links $\sigma_1$ $(a)(0)$, $a = 1, \ldots, r_1$, that couple some pairs of nodes, and a coboundary operator $\Delta_0$ that makes $\mathbf{s}_0 = \Delta_0 \mathbf{x}_0$.

Now suppose that the initial network $\mathcal{L}_1$ (0) is deformed to another network of points $\mathcal{L}_1$ (1) that is associated with another position-displacement complex $N$ (1) = $\{O, \mathbf{x}_1, \mathbf{s}_1, \Delta_1\}$ in such a way that each node $\sigma_0$ $(i)(0)$ of $\mathcal{L}_1$ (0) will go to some corresponding node $\sigma_0$ $(i)(1)$ of $\mathcal{L}_1$ (1), and each link $\sigma_1$ $(a)(0)$ of $\mathcal{L}_1$ (0) will go to some corresponding link $\sigma_1(a)(0)$ of $\mathcal{L}_1$ (1). One can associate each $\sigma_0$ $(i)(1)$ with an absolute position $\mathbf{x}_1$ $(i)$ [relative to the same $O$ that one used to define $\mathbf{x}_0$ $(i)$] and each $\sigma_1(a)(1)$ with a displacement $\mathbf{s}_1$ $(i)$ in such a way that if $\mathbf{s}_0(a) = \mathbf{x}_0$ $(i) - \mathbf{x}_0$ $(j)$ then:

$$\mathbf{s}_1 \ (a) = \mathbf{x}_1 \ (i) - \mathbf{x}_1 \ (j). \tag{6.1}$$

This has the effect of making:

$$\mathbf{s}_1 = \Delta_1 \mathbf{x}_1 ; \tag{6.2}$$

i.e., $f$ is a chain map:

$$\Delta_1 \cdot f = f \cdot \Delta_0 . \tag{6.3}$$

The simplest possibility for this situation is what we shall call an *elementary deformation* of the network, which is characterized by saying that the chain map $f : N_0 \rightarrow$ $N_1$ is a bijection. The next most elementary possibility is that some of the nodes of $N_0$ go to the same node of $N_1$, and to be consistent with the coboundary operator the corresponding displacement between those nodes of $N_0$ would have to go to zero. Finally, if one wished to deal with "topology-changing processes," for which the function $f$ might be only a relation, then one would have to drop the restriction that the map should commute with the coboundary operators.

We next add a set of links $K_1 = \{ \bar{\sigma}_1(i) , i = 1, \ldots, r_0 \}$ such that:

$$\partial \bar{\sigma}_1(i) = \sigma_0(i)(1) - \sigma_0(i)(0) , \qquad (6.4)$$

and extend the set of 1-cosimplexes by their reciprocals $\bar{\sigma}^1(i)$ accordingly.

An elementary deformation of a network $N_0$ can then be associated with a *displacement vector field* **u**, which is a 1-cochain:

$$\mathbf{u} = \sum_{i=1}^{r_0} \mathbf{u}(i) \bar{\sigma}^1(i) , \qquad \mathbf{u}(i) = \mathbf{x}_1(i) - \mathbf{x}_0(i) \equiv \overline{\Delta \mathbf{x}}(i) , \qquad (6.5)$$

with:

$$\overline{\mathbf{x}}(i) \equiv \mathbf{x}_0(i) + \mathbf{x}_1(i), \qquad (6.6)$$

that describes the displacement of each node of $\mathcal{L}_1(0)$ to a corresponding node of $\mathcal{L}_1(1)$ .

We illustrate this situation in Fig. 4:

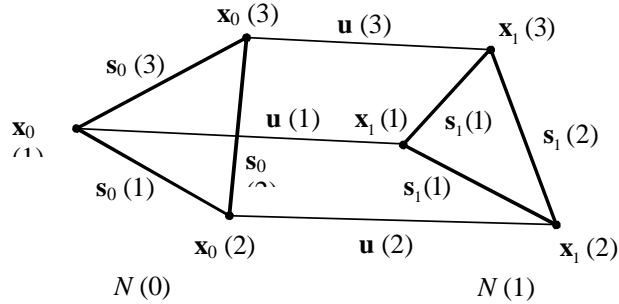

Figure 4. An elementary deformation of a network of points.

A glance at the picture shows that the collective effect of the elementary motion is to define a larger network of points $\mathcal{L}_1 = \mathcal{L}_1(0) \cup \mathcal{L}_1(1) \cup K_1$ whose coboundary operator $\overline{\overline{\Delta}}$ is the formal sum of $\Delta_0$, $\Delta_1$, and $\overline{\Delta}$, such that:

$$\overline{\overline{\Delta}} \, \overline{\mathbf{x}} \equiv (\Delta_0 + \Delta_1 + \overline{\Delta}) \, \overline{\mathbf{x}} = \Delta_0 \, \mathbf{x}_0 + \Delta_1 \, \mathbf{x}_1 + \mathbf{u} . \qquad (6.7)$$

Indeed, if one thinks of the rectangular "faces" of the polygon that are defined by the $\bar{\sigma}_1(i)$ as being 2-dimensional simplexes $\bar{\sigma}_2(i)$, $i = 1, 2, 3$ then the boundary of the 2-chain:

$$\overline{c}_2 = \bar{\sigma}_2(1) + \bar{\sigma}_2(2) + \bar{\sigma}_2(3) \qquad (6.8)$$

will be:

$$\overline{\partial}\,\overline{c}_2 = z_1\,(1) - z_1\,(0)\,, \tag{6.9}$$

with

$$z_1\,(0) = \sigma_1\,(1)(0) + \sigma_1\,(2)\,)(0) + \sigma_1\,(3)(0), \tag{6.10}$$

$$z_1\,(1) = \sigma_1\,(1)(1) + \sigma_1\,(2)\,)(1) + \sigma_1\,(3)(1). \tag{6.11}$$

Hence, the two 1-cycles $z_1\,(0)$ and $z_1\,(1)$ that represent the 1-skeletons of the structural complexes $\mathcal{L}_1\,(0)$ and $\mathcal{L}_1\,(1)$, respectively, will be homologous. Indeed, since we have associated the 2-simplexes $\overline{\sigma}_2(i)$ with a vector field $\mathbf{u}$ that connects corresponding vertices of the two networks, we can call it a *vector homology*. That displacement vector field $\mathbf{u}$ will then represent a motion in time, whereas the initial and final displacement vector fields $\mathbf{s}_0$ and $\mathbf{s}_1$, resp., will represent only static spatial relationships between the vertices of those structures.

One can also start with a vector homology between the networks as the definition of a deformation. That will then include the topology-changing processes, since a vector homology might take a network of points to a single point, which would be essentially a contraction, or take a path-connected network to a network with more than one path component.

In this example, what we will call the *kinematical complex* $\mathcal{K} = \{K_0\,, K_1\,, \overline{\partial}\,\}$ of the deformation consists of the set of 0-simplexes $K_0 = \{\sigma_0\,(i)(A), i = 1, 2, 3, A = 0, 1\}$, the set of 1-simplexes $K_1 = \{\overline{\sigma}_i(i), i = 1, 2, 3\}$, and the boundary operator $\overline{\partial}$ that connects them. We could also include the 2-simplexes that arise from connecting the initial structural links with the final ones, but we shall not actually need them in what follows. The position-displacement complex $\{O, \mathbf{x}_0\,, \mathbf{x}_1\,, \mathbf{u}, \overline{\Delta}\,\}$ that is associated with $\mathcal{K}$ will then consist of the two position 0-cochains $\mathbf{x}_0$ and $\mathbf{x}_1$, the displacement vector field $\mathbf{u}$, and the coboundary operator $\overline{\Delta}$ that connects them.

Another glance at the last figure will show that the lines that represent successive stages of the deformation of an initial mechanical network have been drawn thicker than the ones that connect the vertices of those two networks. The thicker lines are intended to single out the structural links from the thinner kinematical ones.

*b. Concatenation of deformations.* – So far, we have defined a linear displacement of each initial vertex over just a single time interval. There is nothing to stop us from continuing that process recursively over a finite sequence of time points $t_0, \ldots, t_N$, which will be referred to by their subscripts $A = 0, \ldots, N$. In particular, each initial vertex $\sigma_0$ $(i)(0)$ will go to a succession of later vertices $\sigma_0\,(i)(A)$, $A = 1, \ldots, N$. The 1-simplexes $\overline{\sigma}_1(i)(A)$ that connect each pair of successive vertices in the time sequence will then define a path for each $i$. In some cases, the path might return to the starting vertex, in which case the path will define a loop, which is an elementary 1-cycle. If all of the initial vertices lie on loops then we shall call the vector homology *completely cyclic*. In such a case the initial network of points will return to its initial state.

However, there is a subtlety to be addressed here: Although the final vertex $\sigma_0\,(i)(N)$ will coincide with the initial vertex $\sigma_0\,(i)(0)$ in space, nonetheless, if one introduces time as one of the dimensions of the space of motion then one must realize that those two points do not overlap in *space-time*. Similarly, the path does not close to a loop in space-time. Hence, we can speak of kinematical loops and 1-cycles only in regard to the projection of the space-time complex onto space, which we shall call the *spatial trace* of the motion.

Perhaps the simplest example of that situation is given by circular motion in a plane. When one adds time as a third dimension, one will see that the circular motion in the plane will become a helix in space-time. However, the helix is topologically equivalent to the real line, while the circle is not.

*c. The kinematical state of a mechanical complex.* – If one allows each (nodal) point $x(i)(0)$ of in initial structural complex $\mathcal{L}_1(0)$ to move smoothly in time then each node $i$ will describe a curve $x(i)(t)$ in space that does not have to be a straight line segment. However, since we are dealing with discrete systems of points here, we will be associating the kinematical vectors with successive positions of each vertex and the 1-simplexes that connect them.

By differentiating $x(i)(t)$ with respect to $t$ and evaluating that derivative at $t = 0$, one can associate an initial *absolute velocity* vector $\mathbf{v}(i)(0)$ with each node $x(i)(0)$, and therefore an initial *absolute velocity* 0-cochain $\mathbf{v}(0)$ with values in $\mathbb{R}^n$ :

$$\mathbf{v}(0) = \sum_{i=1}^{r_0} \mathbf{v}(i)(0)\,\sigma^0(i)(0) \tag{6.12}$$

that it is consistent with the initial position 0-cochain:

$$\mathbf{x}(0) = \sum_{i=1}^{r_0} \mathbf{x}(i)(0)\,\sigma^0(i)(0)\,, \tag{6.13}$$

in the sense that:

$$\mathbf{v}(0) = \frac{d\mathbf{x}}{dt}\bigg|_{t=t_0} = \sum_{i=1}^{r_0} \dot{\mathbf{x}}(i)(0)\,\sigma^0(i)(0)\,. \tag{6.14}$$

We can define the initial *relative velocity* 1-cochain $\dot{\mathbf{s}}(0)$ with values in $\mathbb{R}^n$ by differentiating the initial displacement $\mathbf{s}(0)$, as well:

$$\mathbf{s}(0) = \sum_{a=1}^{r_1} \mathbf{s}(a)(0)\,\sigma^1(a)(0)\,, \qquad\qquad \dot{\mathbf{s}}(0) = \frac{d\mathbf{s}}{dt}\bigg|_{t=t_0} = \sum_{a=1}^{r_1} \dot{\mathbf{s}}(a)(0)\,\sigma^1(a)(0)\,. \tag{6.15}$$

Since the operator $d/dt$ commutes with $\Delta$, differentiating (5.10) will give:

$$\dot{\mathbf{s}}(0) = \Delta\dot{\mathbf{x}}(0) = \Delta\mathbf{v}(0), \tag{6.16}$$

when (6.14) is valid; i.e., the relative velocity of two moving points will be the time derivative of the displacement vector between them.

A time derivation of the velocity 0-cochain will produce the initial *absolute acceleration* 0-cochain:

$$\mathbf{a}\,(0) = \left.\frac{d\mathbf{v}}{dt}\right|_{t=t_0} = \sum_{i=1}^{r_0} \dot{\mathbf{v}}(i)(0)\,\sigma^0(i)(0)\,, \qquad (6.17)$$

and when (6.14) is valid, this will also give:

$$\mathbf{a}\,(0) = \sum_{i=1}^{r_0} \ddot{\mathbf{x}}(i)(0)\,\sigma^0(i)(0)\,. \qquad (6.18)$$

One will then get the initial *relative acceleration* 1-cochain $\ddot{\mathbf{s}}\,(0)$ from the relative velocity 1-cochain analogously:

$$\ddot{\mathbf{s}}\,(0) = \left.\frac{d\dot{\mathbf{s}}}{dt}\right|_{t=t_0} = \sum_{a=1}^{r_1} \ddot{\mathbf{s}}\,(a)(0)\,\sigma^1(a)(0)\,, \qquad (6.19)$$

along with a relationship between $\ddot{\mathbf{s}}\,(0)$ and $\mathbf{a}\,(0)$ that corresponds to (6.16):

$$\ddot{\mathbf{s}}\,(0) = \Delta\dot{\mathbf{v}}(0) = \Delta\mathbf{a}\,(0), \qquad (6.20)$$

when (6.16) is valid.

That process can be extended to derivatives of indefinitely-high order, but we shall have no need for any of the remaining expressions.

Although we have made all of these definitions for the initial state at $t_0$, they can be easily extended to a sequence $t_A$, $A = 0, \ldots, N$ of more than one time point by summing over all $A$. For example, we can define:

$$\mathbf{x} = \sum_{A=0}^{N} \mathbf{x}\,(A)\,, \quad \mathbf{v} = \sum_{A=0}^{N} \mathbf{v}\,(A)\,, \quad \mathbf{a} = \sum_{A=0}^{N} \mathbf{a}\,(A)\,, \quad \ldots \qquad (6.21)$$

$$\mathbf{s} = \sum_{A=0}^{N} \mathbf{s}\,(A)\,, \quad \dot{\mathbf{s}} = \sum_{A=0}^{N} \dot{\mathbf{s}}\,(A)\,, \quad \ddot{\mathbf{s}} = \sum_{A=0}^{N} \ddot{\mathbf{s}}\,(A)\,, \quad \ldots \qquad (6.22)$$

We shall regard the *absolute kinematical state* of the network of points in space at time $t_A$ as being the ordered $3n+1$-tuple $X_{\text{abs}}\,(t_A) \equiv (t_A, \mathbf{x}\,(A), \mathbf{v}\,(A), \mathbf{a}\,(A))$ and its *relative kinematical state as* $X_{\text{rel}}\,(t_A) \equiv (t_A, \mathbf{s}\,(A), \dot{\mathbf{s}}(A), \ddot{\mathbf{s}}(A))$. Hence, if we denote the $(3n+1)$-dimensional vector space in which those two cochains take their values to be $J^2\,(\mathbb{R}, \mathbb{R}^n)$ then we can think of the absolute kinematical state of the network of points to be an element $X_{\text{abs}}\,(t_A) \in C^0\,(\mathcal{L}_1\,;\,J^2(\mathbb{R}, \mathbb{R}^n))$, while the relative kinematical state is an element $X_{\text{rel}}\,(t_A) \in C^1\,(\mathcal{L}_1\,;\,J^2(\mathbb{R}, \mathbb{R}^n))\,.$

When one combines equations (5.10), (6.16), and (6.20), one will see that one can extend the coboundary operator $\Delta$ to a linear map $\Delta : C^0 (\mathcal{L}_1 \,;\, J^2 (\mathbb{R}, \mathbb{R}^n)) \rightarrow C^1 (\mathcal{L}_1 \,;\, J^2 (\mathbb{R}, \mathbb{R}^n))$ that takes the absolute kinematical state $X_{\text{abs}} (t_A)$ to the relative kinematical state:

$$X_{\text{rel}} (t_A) = \Delta X_{\text{abs}} (t_A) = (t_A, \Delta \mathbf{x} (A), \Delta \mathbf{v} (A), \Delta \mathbf{a} (A)) \,. \tag{6.23}$$

From now on, we shall drop the explicit reference to $t_A$, for brevity.

When the position-displacement complex has a distinguished point $O$, one can then think of all the absolute kinematical variables at each node as being defined with respect to $O$, as well. However, although one usually sets $\mathbf{x} (O) = 0$, it is not necessary to require that $\mathbf{v} (O) = 0$ and $\mathbf{a} (O) = 0$. One has:

$$\mathbf{v} (i) = \dot{\mathbf{r}} (O, i) \,, \qquad \mathbf{a} (i) = \ddot{\mathbf{r}} (O, i) \,, \tag{6.24}$$

at each node $i \neq O$.

Under a change of $O$ to $O' = O + \mathbf{b} (t)$, in addition to the shifts in the absolute positions $\mathbf{x} (i)$ to $\mathbf{x}' (i)$, with $\mathbf{x}' (i) - \mathbf{x} (i) = - \mathbf{b} (t)$, one must also have corresponding shifts in the absolute velocities and accelerations that are obtained by differentiating that relationship [1]:

$$\mathbf{v}' (i) - \mathbf{v} (i) = - \dot{\mathbf{b}} \,, \qquad \mathbf{a}' (i) - \mathbf{a} (i) = - \ddot{\mathbf{b}} \,. \tag{6.25}$$

However, the relative kinematical state will be unaffected by the shift in reference point.

Hence, we can now think of the change in the kinematical state $(t_A, - \mathbf{u}, - \dot{\mathbf{u}}, - \ddot{\mathbf{u}})$ as being associated with a 0-cochain that associates each node at $t_A$ with the same value. According to (5.18), the 0-chain that is defined by $- \mathbf{b}$ is a 0-cocycle, so differentiating that formula will yield that the 0-cochain $Z^0 (t_A)$ that assigns the constant value $(t_A, - \mathbf{u}, - \dot{\mathbf{u}}, - \ddot{\mathbf{u}})$ to every node will be a 0-cocyle with values in $J^2 (\mathbb{R}, \mathbb{R}^n)$; i.e.:

$$\Delta Z^0 (t_A) = 0. \tag{6.26}$$

Since that means:

$$X'_{\text{abs}} (t_A) - X_{\text{abs}} (t_A) = Z^0 (t_A) = (\text{0-cocycle}), \tag{6.27}$$

we can extend our previous analogy of gauge transformations of the position complex to gauge transformations of the kinematical states.

Eventually, one must also deal with the fact that many of the possible abstractly-given states of the forms $(t_A, \mathbf{x} (A), \mathbf{v} (A), \mathbf{a} (A))$ do not have the form $(t_A, \mathbf{x} (A), \dot{\mathbf{x}} (A), \ddot{\mathbf{x}} (A))$, but only the "integrable" ones. In fact, the non-integrable case has considerable physical significance, due to the fact that it is what one gets from using anholonomic frame fields for the description of the components of $\mathbf{v}$ and $\mathbf{a}$. However, addressing the question of integrability would take us far afield from what we have left to do, which is already quite substantial, so we will let it pass with only that brief mention.

---

[1] Note this is all very non-relativistic, since the relativistic formula for the addition of velocities is more involved that this, and the relativity of acceleration is still open to debate.

*d. Rigid motions and deformations of structural complexes.* – We shall define an *infinitesimal motion* of a structural complex at some point in time to be a vector field $\mathbf{v}(i)$ that is defined on the nodes and represents the velocity of each node, or rather, the infinitesimal generator of a one-parameter family of finite motions. Depending upon how the nodes of the structural complex are linked, there will be two basic types of motions in space that the network can exhibit: Rigid motions and non-rigid motions or deformations. The difference between them leads one into the deeper issue of how to define the "rigidity" of a network.

A *rigid motion* of a network in space is basically described by a vector field on the nodes with the property that all of the distances between the nodes (using whatever metric is defined on the ambient space) will remain constant under the motion. In particular, note that this does not require the nodes to be connected by a link.

If there are $r_0$ nodes then there will be $(r_0)^2$ pairs of them, although since the basic properties of distance include the idea that the distance between any two nodes is symmetric, one can reduce the Cartesian product $S_0 \times S_0$ of all ordered pairs $(\sigma_0(i), \sigma_0(j))$ to the set $\Sigma$ of all unordered pairs $\{\sigma_0(i), \sigma_0(j)\}$, which we will abbreviate to simply $\{i, j\}$. The set $\Sigma$ will then have $r_0(r_0 +1) / 2$ elements. Hence, the distance function will take the form of $\|\cdot\| : \Sigma \to \mathbb{R}, \{i, j\} \mapsto \|\{i, j\}\|$, which also defines a vector in $\mathbb{R}^{r_0(r_0+1)/2}$, namely, $(\|\{1,1\}\|, \|\{1, 2\}\|, \ldots, \|\{r_0, r_0\}\|)$, which represents the ordered set of all distances between nodes, and includes the fact that the distance from any node to itself must be zero: $\|\{i,i\}\| = 0$. If one allows each node $\sigma_0(i)$ to vary in time $\sigma_0(i)(t)$ then so will all of the distances, in general: $\|\{i, j\}\|(t)$; that will define a curve in $\mathbb{R}^{r_0(r_0+1)/2}$. The motion of the network is *rigid* if the distances are all constant in time, and thus define a constant function from $\Sigma$ to $\mathbb{R}^{r_0(r_0+1)/2}$; i.e., a constant vector in $\mathbb{R}^{r_0(r_0+1)/2}$.

When the metric that $\mathbb{R}^n$ is given is the Euclidian one, there will be two types of rigid motions to consider: translations and rotations. A translation will be represented by a constant vector field on the nodes, which will then define a 0-cocyle as in (5.15).

A rotation is somewhat more involved, since one always has to specify the point (in the plane) or the axis (in space) about which the rotation takes place. When one starts with points in an affine space $A^n$, this comes down to the choice of origin $O$ and the associated map $s_O : A^n \to \mathbb{R}^n$ that models the affine space on a vector space. For any choice of $O$, the rotation can be represented by a rotation $R$ of vectors in $\mathbb{R}^n$ about an axis through its natural origin 0 and its $n \times n$ rotation matrix $R^i_j$. The axis will be generated by an eigenvector with an eigenvalue of 1 when $n = 3$, and when $n = 2$, the rotation will take place about 0. For $n = 1$, the only non-trivial rotation is multiplication by $-1$, although that is actually an improper rotation. The difference between proper and improper rotations is that a proper rotation can be connected to the identity transformation by a continuous curve, while an improper one cannot. Typically, an improper rotation is the composition of a rotation with a reflection through 0, a line, or a plane, in the cases of rotations in a line, a plane, and space, resp.

In order to get a velocity vector field on $\mathbb{R}^n$ out of a rotation $R$, one defines the *fundamental vector field* for $R$ under the action of the rotation group on $\mathbb{R}^n$. If $R$ takes the form:

$$R = \exp \omega = \sum_{n=0}^{\infty} \frac{1}{n!} \omega^n, \qquad (0! = 1, \ \omega^0 = I) \qquad (6.28)$$

for some antisymmetric matrix $\omega$, which can happen only for proper rotations, then one can think of $R$ as the value $R(0)$ for a differentiable one-parameter family (subgroup, in fact) of rotations:

$$R(t) = \exp(\omega t). \qquad (6.29)$$

When $R(t)$ acts upon $\mathbb{R}^n$, it will take any vector $\mathbf{x} \in \mathbb{R}^n$ to a one-parameter family of vectors:

$$\mathbf{x}(t) = R(t)\,\mathbf{x}, \qquad (6.30)$$

which will define a unique curve (viz., a circle about the origin) through each $\mathbf{x}$.

The velocity vector field that one gets from this is then the velocity vector that is associated with each point on each curve, namely:

$$\mathbf{v} = \frac{d}{dt}\bigg|_{t=t_0} \mathbf{x}(t) = \frac{d}{dt}\bigg|_{t=t_0} R(t)\,\mathbf{x} = \omega\,\mathbf{x}. \qquad (6.31)$$

The matrix $\omega$, which is the infinitesimal generator of the finite rotation $R$, then takes the form of the *(orbital) angular velocity* of the family of rotations $R(t)$ at $t = 0$. One can then map the velocity vectors $\mathbf{v}$ in $\mathbb{R}^n$ back to vectors that are tangent to the nodes of the network by inverting the map $s_O$ and differentiating it.

Sometimes a judicious choice of $O$ can simplify the description of motion. For planar rigid motions, one can look for an "instantaneous point of rotation" that might make all of the nodal velocity vectors take the form of a fundamental vector field for some rotation. For spatial rigid motions, one can look for an "instantaneous axis of rotation." Hence, the decomposition of a rigid motion in a metric affine space into a proper rotation and a translation is not unique, in general. However, a rigid motion in space does admit a unique canonical representation as a rotation about a "central axis" and a translation along it, which is referred to as a "screw." (See, e.g., [**19**] for a discussion of that.)

Note that if one had defined rigidity to be constancy of the lengths of the links then one would be including some motions that allowed distances between the unlinked nodes to change nonetheless. That is how the issue of the rigidity of mechanical networks (e.g., frameworks) is more subtle that it sounds. There can be such things as *internal degrees of freedom*, which are, in a sense, dual to internal stresses. For instance, the pin-jointed triangle is a rigid network in that it has no internal degrees of freedom, while the pin-jointed rectangle is not rigid, since it can be deformed by a shearing motion. These possibilities are illustrated in Fig. 5 below.

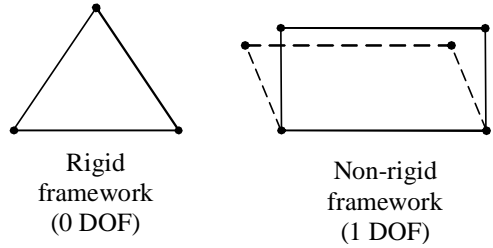

Rigid
framework
(0 DOF)

Non-rigid
framework
(1 DOF)

Figure 5. – Rigid and non-rigid frameworks.

One can see that although the lengths of the sides of the rectangle do not change under the deformation, it is nonetheless not a rigid motion. One says that it has one internal degree of freedom that comes from the freedom in the vertex angles to change, but not independently. Since a link can translate in space, as well as rotate, under such an internal motion, one would be justified in calling it a *system of internal rigid motions*. The total number of (non-rigid) degrees of freedom in a network can then be defined by *Maxwell's formula* [**20**]:

$$\text{DOF} = n \, r_0 - r_1 - n \, (n+1) \, / \, 2 = \begin{cases} 2r_0 - r_1 - 3 & n = 2, \\ 3r_0 - r_1 - 6 & n = 3. \end{cases} \qquad (6.32)$$

This formula is easy to justify: $r_0$ nodes in an $n$-dimensional space will have $n \, r_0$ degrees of freedom in the absence of any couplings between them. Every rigid link will subtract one degree of freedom from that total, and there are $r_1$ links. The last expression $n \, (n + 1) \, / \, 2$ describes the degrees of freedom that come from a collective rigid motion of the network. For instance, in the two planar figures that are depicted in Fig. 5, the triangle has 3 nodes and 3 links, so it has 2 (3) − 3 − 3 = 0 degrees of freedom, while the rectangle has 4 nodes and four links, so it will have 2 (4) − 4 − 3 = 1 degree of freedom by this method of accounting.

A final possibility for the motion of a mechanical network is that in addition to an overall rigid motion and other motions that preserve the lengths of the links, one might also imagine motions that change the lengths of the links, if they are assumed to be flexible, in their own right. In such a case, one will be dealing with a general *deformation* of the framework. Typically, one subtracts out the part of the nodal velocity distribution that can be attributed to an overall rigid motion in order to concentrate on just the pure deformations.

**7. Dynamics. −** Dynamical states are dual to kinematical states (or rather, to *infinitesimal perturbations* of the kinematical states) in the sense that there is a natural bilinear pairing of the two that produces a real number whose physical interpretation is that of virtual work, which we shall return to in the next section. We shall first discuss how linear momentum is dual to velocity, and then examine the duality of force and displacement in the following section.

Dynamical states also embody the empirical nature of the matter that is involved in a mechanics problem, so there is generally no way of making them purely geometrical constructions like kinematical states. The association of a dynamical state with a virtual displacement of a kinematical state always involves the definition of a *mechanical constitutive law.*

We shall now attempt to clarify the details of that situation in the language of homology.

*a. The mass 0-chain.* Suppose that we are given a structural complex $\mathcal{L}_1$ at some initial time point $t_0$ . We basically start by assuming that each (initial) node $\sigma_0(i)$ is associated with a positive scalar $m(i)$ that represents the (initial) mass of that point. That suggests that we define the 0-chain with real coefficients:

$$m = \sum_{i=1}^{r_0} m(i)\,\sigma_0(i)\,, \tag{7.1}$$

which we can think of as the *total mass* 0-chain of the network of points; i.e., $m \in C_0(\mathcal{L}_1;$ $\mathbb{R})$. We can see that:

$$M = \partial^\# m \equiv \sum_{i=1}^{r_0} m(i) \tag{7.2}$$

is, in fact, the total mass when we go to "extended" homology.

We can make the masses vary in time differentiably (to some order) and evaluate them at successive time points $t_A, A = 0, \ldots, N$ :

$$m(A) = \sum_{i=1}^{r_0} m(i)(A)\,\sigma_0(i)(A)\,, \tag{7.3}$$

and then define successive derivatives of the total mass chain at each time point. We shall only concern ourselves with the first one:

$$\dot{m}(A) = \frac{dm}{dt}\bigg|_{t=t_A} = \sum_{i=1}^{r_0} \dot{m}(i)(A)\,\sigma_0(i)(A)\,, \tag{7.4}$$

which is then a 0-chain with coefficients in $C^\infty(\mathbb{R})$ ; i.e., $\dot{m}(A) \in C_0(\mathcal{L}_1\,;\,C^\infty(\mathbb{R}))$.

One can now associate each 1-simplex $\sigma_1(a)(A)$ in the network at time $t_A$ with a real-valued function $J_a(A)$ that we shall call the *mass flow rate* through that branch at $t_A$ and form the 1-chain with coefficients in $C^\infty(\mathbb{R})$:

$$J_1(A) = \sum_{a=1}^{r_1} J(a)(A)\,\sigma_1(a)(A)\,. \tag{7.5}$$

Hence, $J_1(A) \in C_1(\mathcal{L}_1 ; C^\infty(\mathbb{R}))$.

If we take the boundary of this:

$$\partial J_1(A) = \sum_{a=1}^{r_1} J(a)(A)\,\partial\sigma_1(a)(A) = \sum_{i=1}^{r_0} \left[\sum_{a=1}^{r_1} J_a(A)\,\partial_a^i\right]\sigma_0(i)(A) \tag{7.6}$$

then we will see that its coefficients each represent the signed sum of all the mass flow rates that are incident on that node. Hence, we then pose the:

**Balance of mass axiom**: *The time derivative of the mass at each node equals the sum of the mass flow rates that are incident on that node:*

$$\dot{m}_i(A) = \sum_{a=1}^{r_1} J(a)(A)\,\partial_a^i\,, \tag{7.7}$$

hence:

**Balance of mass theorem:**
$$\dot{m}_i(A) = \partial J_1(A). \tag{7.8}$$

*That is: the time derivative of the total mass chain is a boundary.*

Furthermore:

**Conservation of mass corollary:**

*The total mass is conserved iff $J_1(A)$ is a 1-cycle.*

Had we been talking about charge, instead of mass, this would correspond to Kirchhoff's law of currents, which is, of course, based upon the conservation of charge.

*b. The mass moment 0-chain*. – The constructions of this section will be made for a single time point $t_A$, but their extension to a sequence of time points is immediate.

Once again, one must assume that the network of points contains a distinguished point $O$. Although the only moment that was defined above was the moment of a vector about $O$, one can also define the moment of a scalar analogously by regarding scalar multiplication of 1-form as exterior multiplication by a 0-form. However, this time, it is the position covector field $\{r(i), i = 1, \ldots, r_0\}$ of every nodal point with respect to $O$ that will be used, since we will be eventually treating linear momentum as a 1-form, rather than a vector. The position 0-cochain is then defined by:

$$r = \sum_{i=1}^{r_0} r(i)\,\sigma_0(i)\,. \tag{7.9}$$

in which:

$$r(i) = x_i(i)\,dx^i\,, \tag{7.10}$$

and $r$, $r(i)$, and the $x_i(i)$ are all smooth functions of time.

One can combine the mass 0-chain with the components of the position 0-cochain and get the *mass moment* (with respect to $O$) 0-chain, which has its coefficients in the vector space $C^\infty(\mathbb{R}; \mathbb{R}^{n^*})$, this time:

$$\mathbf{M}[O] = \sum_{i=1}^{r_0} m(i)\,r(i)\,\sigma_0(i)\,. \tag{7.11}$$

Hence, the augmented boundary of this 0-chain will be:

$$\partial^\# \mathbf{M}[O] = \sum_{i=1}^{r_0} m(i)\,r(i)\,, \tag{7.12}$$

which is a 1-form that then becomes the *total mass moment about $O$*.

The *center-of-mass* of the total mass 0-chain is defined (at any instant, if the masses are time-varying) to be the point $O = x_{\text{cm}}$ in space with the property that the total mass moment with respect to that point vanishes:

$$\partial^\# \mathbf{M}[x_{\text{cm}}] = 0\,. \tag{7.13}$$

Hence, if $r(i) = x(i) - x_{\text{cm}}$ then one will get:

$$\mathbf{M}[x_{\text{cm}}] = \sum_{i=1}^{r_0} m(i)\,[x(i) - x_{\text{cm}}]\,\sigma_0(i)\,, \tag{7.14}$$

so if one now regards $x(i) - x_{\text{cm}}$ as the difference of two 1-forms then:

$$\begin{aligned}
\partial^\# \mathbf{M}[x_{\text{cm}}] &= \sum_{i=1}^{r_0} m(i)\,[x(i) - x_{\text{cm}}] = \left[\sum_{i=1}^{r_0} m(i)\,x(i)\right] - \left[\sum_{j=1}^{r_0} m(j)\right] x_{\text{cm}} \\
&= \left[\sum_{i=1}^{r_0} m(i)\,x(i)\right] - M\,x_{\text{cm}}\,,
\end{aligned}$$

and if this equals 0 then one must have:

$$M\,x_{\text{cm}} = \sum_{i=1}^{r_0} m(i)\,x(i) \qquad \text{or} \qquad x_{\text{cm}} = \sum_{i=1}^{r_0} \frac{m(i)}{M}\,x(i)\,. \tag{7.15}$$

If one regards $m(i)/M$ as a probability density function, instead of a mass percentage, then the center-of-mass will represent essentially a "mean position" of the points in the network, as weighted by the masses.

When $O$ goes to $O' = O + \mathbf{b}$, the position covector field $r(i)$ will change to $r'(i) = r(i) - b$, where $b$ is the covector that is dual to the vector $\mathbf{b}$, and the new mass moment about $O'$ will be:

$$\mathbf{M}[O'] = \sum_{i=1}^{r_0} m(i)\, r'(i)\, \sigma_0(i) \;=\; \mathbf{M}[O] - \sum_{i=1}^{r_0} m(i)\, b\, \sigma_0(i) \,. \qquad (7.16)$$

Thus, if $O$ is arbitrary and $O' = x_{\text{cm}}$ is the center of mass then one can characterize the displacement $\mathbf{b}$ between them by the equality:

$$\partial^{\#}\mathbf{M}[O] = \sum_{i=1}^{r_0} m(i)\, r(i) \;=\; \sum_{i=1}^{r_0} m(i)\, b = M\, b \;;$$

which will make:

$$b = \sum_{i=1}^{r_0} \frac{m(i)}{M}\, r(i) \;=\; x_{\text{cm}} - O \,. \qquad (7.17)$$

*c. Linear momentum.* If one differentiates (7.11) with respect to time then one will get:

$$\dot{\mathbf{M}}[O] = \sum_{i=1}^{r_0} \dot{m}(i)\, r(i)\, \sigma_0(i) + \sum_{i=1}^{r_0} m(i)\, \dot{r}(i)\, \sigma_0(i) \,. \qquad (7.18)$$

The first summation can be regarded as the formal sum of the moments of the various mass flow rates at each node with respect to the origin, while the second one:

$$p[O] = \sum_{i=1}^{r_0} p(i)\, \sigma_0(i) \,, \qquad p(i) \equiv m(i)\, \dot{r}(i) \qquad (7.19)$$

can be called the *total linear momentum 0-chain* (relative to $O$) of the mass network, and it takes its coefficients from the vector space $C^{\infty}(\mathbb{R}; \mathbb{R}^{n*})$. Its augmented boundary will be the 1-form:

$$\partial^{\#} p[O] = \sum_{i=1}^{r_0} p(i) \,, \qquad (7.20)$$

which will then be the *total linear momentum of the network (relative to $O$)*.

One then has:

**Theorem:**

*When the masses that are associated with each node are constant in time, one will have:*

$$p\,[O] = \dot{\mathbf{M}}[O]\,. \tag{7.21}$$

*i.e.: linear momentum is the time derivative of the mass moment.*

When one changes from $O$ to $O' = O + \mathbf{b}$, $\dot{r}(i) = \dot{x}(i) - \dot{x}(O)$ will go to:

$$\dot{r}'(i) = \dot{x}(i) - \dot{x}(O') = \dot{x}(i) - \dot{x}(O) - \dot{b} = \dot{r}(i) - \dot{b}\,, \tag{7.22}$$

and $p\,(i)$ will go to:

$$p\,'(i) = p\,(i) - m\,(i)\,\dot{b}\,, \tag{7.23}$$

so:

$$p\,[O'] = p\,[O] - \sum_{i=1}^{r_0} [m(i)\dot{b}]\sigma_0(i)\,, \tag{7.24}$$

and $\partial^{\#}p\,[O]$ will go to:

$$\partial^{\#}p\,[O'] = \partial^{\#}p\,[O] - \sum_{i=1}^{r_0} m(i)\ddot{b} = \partial^{\#}p\,[O] - M\,\dot{b}\,. \tag{7.25}$$

Recall the discussion above of the force complex of a mechanical network.

**Axiom (Newton's third law, for forces):**

*When one inverts the orientation of $\sigma_1\,(a)$, one must invert the sign of $\mathbf{F}\,(a)$ in order to compensate.*

We then pose the:

**Balance of linear momentum axiom (Newton's second law):**

*The resultant force at each node $i$ of the mechanical network equals the time derivative of the momentum of that mass:*

$$F_{\text{res}}\,(i)\ = \frac{dp\,(i)}{dt}\,, \qquad\qquad F_{\text{res}} \equiv F_{\text{ext}} + \partial F_{\text{int}}\,. \tag{7.26}$$

That is:

**Balance of linear momentum theorem:**

$$\frac{dp\,[O]}{dt} = F_{\text{res}}\,. \tag{7.27}$$

The theorem also takes the form of a generalization of Kirchhoff's law of currents when those currents take their values in $\mathbb{R}^{n*}$.

One can also say:

**Corollary:**

i)   *dp [O] / dt is homologous to $F_{\text{ext}}$ .*

ii)   *One has:*

$$\partial^{\#}\left(\frac{dp[O]}{dt}\right) = \partial^{\#} F_{\text{ext}} \; ; \tag{7.28}$$

*i.e. : the time derivative of the total linear momentum of the network equals the resultant of the external forces that act upon its nodes.*

Hence, the internal forces have no effect on the collective motion of the network.

**Corollary:**

*In the absence of external forces, the time derivative of the total momentum (relative to O) is a boundary; namely, $\partial F_{\text{int}}$ .*

This last possibility relates to only deformable networks in which all of the forces that act upon the nodes are internal. For instance, imagine putting an elastic diagonal link into a pin-jointed rectangle that applies a force of tension to those nodes. All of the forces in the network are internal, but they will still result in motion of the nodes.

We shall now introduce a restricted definition of the total impulse delivered by a time-varying force $F$ $(t)$ that acts along a line segment, which shall mean only the 1-simplexes of a kinematical complex that carry the displacements **u** $(i)$ that represent actual motions of the nodes of an initial structural complex. If we again denote the force that acts along the 1-simplex $\sigma_1$ $(i)$ that carries **u** $(i)$ by $F$ $(i)(t)$ then the *impulse delivered by $F$ $(i)(t)$ along $\sigma_1$ $(i)$* will be:

$$I\,(i)\,[\,\sigma_1\,(i)\,] = \int_{t_0}^{t_1} F(i)(t)\, dt \,. \tag{7.29}$$

Hence, $I$ $(i)$ can be represented by a 1-cochain $I$ $(i)$ $\sigma^1$ $(i)$ with values in $\mathbb{R}^{n^*}$. The sum of all contributions from all of the $F$ $(i)(t)$ that act along links that represent motions will then be the 1-cochain with values in $\mathbb{R}^{n^*}$ :

$$I = \sum_{i=1}^{r_0} I(i)\,\sigma^1(i) \; = \int_{t_0}^{t_1} F(t)\, dt \tag{7.30}$$

which one can regard as the *impulse delivered by the total external forces* ([1]). Its augmented boundary is a 1-form:

---

([1])   Actually, this popular process of integrating vectors by integrating their components is not invariant under any changes of frame except for ones that do not vary in time or space, so its extension to relativistic mechanics has limited significance.

$$\partial^{\#} I = \sum_{i=1}^{r_0} I(i) \qquad (7.31)$$

that can be called the *total impulse delivered to the network by the total external forces*

If one defines the total linear momentum to be the 1-cochain:

$$p = \sum_{i=1}^{r_0} p(i)\sigma^0(i), \qquad (7.32)$$

which still takes its values in $\mathbb{R}^{n^*}$, and then integrates the left-hand side of (7.27) then that will give:

**Theorem (impulse-momentum):**

$$I = \Delta p = p(t_1) - p(t_0). \qquad (7.33)$$

The definition of $\Delta$ and an application of $\partial^{\#}$ will yield:

**Corollary:**

i)   *The total impulse delivered by F is a 1-coboundary.*

ii)  *The total impulse that is delivered to the network (viz., $\partial^{\#}I$) over a time interval is equal to the change in total linear momentum of the network over that time interval.*

Recall that one says that the network is in *force equilibrium* iff $F_{\text{res}} = 0$ iff $F_{\text{ext}} = -\partial F_{\text{int}}$. One then sees that linear momentum is conserved (i.e., constant in time) at the node $i$ iff it is in equilibrium. Hence:

**Corollary (conservation of linear momentum):**

*The following are equivalent:*

1. *The network is in force equilibrium.*
2. *$F_{\text{res}}$ vanishes.*
3. *$F_{\text{ext}}$ is a boundary.*
4. *Momentum is conserved at every node of the network.*

One can also regard this theorem as a generalization of Newton's first law, when it is interpreted in terms of linear momentum. Since it is a corollary to the second law, one cannot regard it as an axiom, which would require it to be independent of the other axioms.

*d. Angular momentum.* – The kind of angular momentum that we shall address is, strictly-speaking, *orbital* angular momentum – i.e., the moment of the linear momentum. However, in order to do justice to the *intrinsic* kind of angular momentum – or "spin" – would take us too far afield for the present purposes.

Suppose that one is given a network $\mathcal{L}_1$ of points in an affine space $A^n$ that is given a metric $g$. If $O$ is a reference point in space then one can form the cone of $\mathcal{L}_1$ by adding the 0-simplex $\sigma_0\,(0)$ that represents the point $O$ and the 1-simplexes $\sigma_1\,(i) = \sigma_1\,(0, i)$ that connect $O$ to the nodes of $\mathcal{L}_1$, so $\partial\sigma_1\,(i) = \sigma_0\,(i) - \sigma_0\,(0)$ for each $i = 1, \ldots, r_0$. One then associates position vector $\mathbf{x}\,(0) = 0$ to $O$ and the displacement vectors $\mathbf{r}\,(i) = \Delta\mathbf{x}\,(i) = \mathbf{x}\,(i) - \mathbf{x}\,(0)$ with the 1-simplexes $\sigma_1\,(i)$. The corresponding position covector $r\,(i)$ for each $i$ will then be defined by "lowering the index" of $\mathbf{r}\,(i)$ using $g$.

Hence, if $p \in C_0\,(\mathcal{L}_1 ; \mathbb{R}^{n^*})$ is the absolute total linear momentum 0-chain relative to some reference point $O$ then each node $i$ of the original network $\mathcal{L}_1$ will be associated with a 2-form $L\,(i) = r\,(i) \wedge p\,(i)$ that one calls the *(orbital) angular momentum of the node i about O*. They can then be assembled in to a 0-chain with coefficients in $\Lambda^2\,\mathbb{R}^n$ :

$$L\,[O] = \sum_{i=1}^{r_0} L(i)\,\sigma_0(i) \qquad\qquad (7.34)$$

that one calls the *total angular momentum 0-chain* with respect to $O$, while its augmented boundary:

$$\partial^{\#}L\,[O] = \sum_{i=1}^{r_0} L(i) \qquad\qquad (7.35)$$

is a 2-form that one might call the *total angular momentum of the network about O*.

Since the definition of the force moment (or torque) follows from associating each $F(i) = F_{\text{ext}}\,(i) + \partial F_{\text{ext}}\,(i)$ with $M\,(i) = r\,(i) \wedge F\,(i)$ and extends to a 0-chain $M$ that also takes its values in $\Lambda^2\,\mathbb{R}^n$, one can take the moment of both sides of Newton's second law (7.27) and get:

$$\sum_{i=1}^{r_0}[r(i) \wedge \dot{p}(i)]\sigma_0(i) \;=\; \sum_{i=1}^{r_0} M(i)\,\sigma_0(i)\,. \qquad\qquad (7.36)$$

In the event that $p\,(i)$ has the convective form $m\,(i)\,\dot{r}\,(i)$ then:

$$r\,(i) \wedge \dot{p}\,(i) \;=\; \frac{dL(i)}{dt} \qquad\qquad \text{for each } i, \qquad\qquad (7.37)$$

which can be combined into the 0-chain $dL\,/\,dt$ with coefficients in $\Lambda^2\,\mathbb{R}^n$. That would make:

**Theorem (balance of angular momentum):**

*When linear momentum is convective, one will have:*

$$\frac{dL}{dt} = M \ . \tag{7.38}$$

Applying the augmented boundary operator to both sides will give:
**Corollary:**

*The time rate of change of the total angular momentum of the network equals the total moment of the external forces that that act upon its nodes.*

Since the moments of internal forces in the original network vanish in pairs, the moment of the forces at each of the nodes consists solely of contributions from the external forces, and the total moment 0-chain $M$ will be a sum of only those moments. One can then say that:

**Corollary (conservation of angular momentum):**

*The following are equivalent:*

1. *The network is in moment equilibrium about O.*
2. *M vanishes.*
3. *Angular momentum is conserved at every node of the network.*

When a force moment $M(i)$ is time-varying, one can define the *moment impulse* that it delivers in a time interval $[t_0, t_1]$ by the integral:

$$I_M(i)[t_0, t_1] = \int_{t_0}^{t_1} M(i)(t)\,dt \ . \tag{7.39}$$

If the motion of the node $i$ during that time interval is associated with the 1-simplex $\sigma_1(i)$, whose reciprocal 1-cosimplex is $\sigma^1(i)$, then those contributions can be assembled into a 1-cochain with values in $\Lambda^2\mathbb{R}^n$ :

$$I_M = \sum_{i=1}^{r_0} I_M(i)\sigma^1(i) \ . \tag{7.40}$$

The augmented boundary of this $I_M$ will be a 2-form:

$$\partial^\# I_M = \sum_{i=1}^{r_0} I_M(i) \tag{7.41}$$

that can be called the *total moment impulse delivered to the network during the time interval* $[t_0, t_1]$.

If we convert the 0-chain $L$ with coefficients in $\Lambda^2 \mathbb{R}^n$ into a 0-cochain by replacing the 0-simplexes with their reciprocal 0-cosimplexes then we will get the following analogue of the impulse-momentum theorem:

**Theorem (moment impulse-angular momentum):**

$$I_M = \Delta L = L(t_1) - L(t_0) . \tag{7.42}$$

and the analogous corresponding:

**Corollary:**

*i) The total moment impulse is a 1-coboundary.*

*ii) The total moment impulse that is delivered to the network over a time interval is equal to the change in total angular momentum of the network over that interval.*

**8. Energetics.** – Energy is a concept that takes on the character of a bilinear pairing of infinitesimal perturbations of kinematical states with dynamical states. There are essentially two types of energy that then relate to a kinematical state of the form $(t, \mathbf{x}(t), \mathbf{v}(t))$, namely, potential energy and kinetic energy. Kinetic energy is a pairing of velocity $\mathbf{v}$ with linear momentum $p$ and can be defined directly from a kinematical state and its associated dynamical state. However, potential energy is associated with the work done by a force over a distance, and unless the force is constant over that distance, the relationship between force, displacement, and work will not be algebraic, but will require integration along the curve, which will be addressed in Part II. In order to deal with work algebraically, one must either consider forces that are constant in time and finite displacements or go to infinitesimal displacements, or "virtual" displacements, which will be addressed in the next section.

Since kinetic energy is associated with an infinitesimal state of motion (viz., velocity), and therefore does not require any integration, we will start with a discussion of that simpler concept.

*a. Kinetic energy.* – Absolute momentum $p$ is a 0-chain with coefficients in $\mathbb{R}^{n^*}$, while absolute velocity $\mathbf{v}$ is a 0-cochain with values in $\mathbb{R}^n$. The bilinear pairing of $p$ with $\mathbf{v}$ is obtained by pairing the 0-simplexes of $p$ with the dual 0-cosimplexes of $\mathbf{v}$ and pairing the covectors of $p$ with the vectors of $\mathbf{v}$:

$$< \mathbf{v}, p > = \sum_{i=1}^{r_0} \sum_{j=1}^{r_0} < p(i), \mathbf{v}(j) > < \sigma^0(j), \sigma_0(i) > = \sum_{i=1}^{r_0} < p(i), \mathbf{v}(i) >, \tag{8.1}$$

which then takes the form of twice the *total (non-relativistic)* ([1]) *kinetic energy* of all moving mass points:

$$\text{KE}_{\text{tot}} = \tfrac{1}{2} < \mathbf{v}, p > . \tag{8.2}$$

One can rearrange the middle equation in (8.1) to take the form:

$$< \text{KE}_{\text{tot}} , \sigma_0(i) > = \tfrac{1}{2} \sum_{j=1}^{r_0} < p(i), \mathbf{v}(j) > < \sigma^0(j), \sigma_0(i) > = \tfrac{1}{2} < p(i), \mathbf{v}(i) > . \tag{8.3}$$

Hence, one can define the real-valued *total kinetic energy 0-cochain* by:

$$\text{KE} = \sum_{i=1}^{r_0} \text{KE}(i)\sigma^0(i), \qquad \text{KE}(i) = \tfrac{1}{2} < p(i), \mathbf{v}(i) > . \tag{8.4}$$

Dually, one can define the *total kinetic energy 0-chain by:*

$$\text{KE} = \sum_{i=1}^{r_0} \text{KE}(i)\sigma_0(i), \tag{8.5}$$

which has the property that its augmented boundary:

$$\partial^{\#} \text{KE} = \sum_{i=1}^{r_0} \text{KE}(i) \tag{8.6}$$

represents the total kinetic energy of the network.

When $\mathcal{L}_1(0)$ and $\mathcal{L}_1(1)$ are two successive states of the network in time, one can define the total kinetic energy 0-cochain for the moved/deformed state $\mathcal{L}_1(1)$ to be:

$$\overline{\text{KE}} = \sum_{i=1}^{r_0} \overline{\text{KE}}(i)\bar{\sigma}^0(i), \tag{8.7}$$

with the analogous definitions for the $\overline{\text{KE}}(i)$.

---

([1])  The extension of the definitions to relativistic kinetic energy first involves replacing each $p(i)$ with an energy-momentum 1-form on Minkowski space, while $\mathbf{v}(i)$ gets replaced with the four-velocity vector $\mathbf{u}(i)$. Since that four-velocity is assumed to be parameterized by proper time, it will have the property that $< \mathbf{u}(i), \mathbf{u}(i) > = c^2$ when one uses the scalar product on Minkowski space. If $p(i)$ takes the form $m_0(i)\, u(i)$, where $u$ is the covelocity 1-form that is dual to $\mathbf{u}$ under that scalar product [i.e., $u(i)(\mathbf{u}(i)) = < \mathbf{u}(i), \mathbf{u}(i) > = c^2$] then:

$$< p(i), \mathbf{u}(i) > = m_0(i)\, c^2.$$

However, the process of adding the various contributions from the nodes will be relativistically proper only when the magnitudes of their relative velocities are very small in comparison to $c$.

If one defines the real numbers $\Delta KE\,(i) \equiv \overline{KE}\,(i) - KE\,(i)$ then one can associate them with the 1-cosimplexes $\sigma^1\,(i)$ and define the 1-cochain:

$$\Delta KE = \sum_{i=1}^{r_0} \Delta KE\,(i)\,\sigma^1(i) = \overline{KE} - KE. \qquad (8.8)$$

Hence, this is a consistent use of the $\Delta$ operator to mean the coboundary, as well as the finite difference.

*b. Work as a 1-cochain.* – Although the discussion of this subsection is perhaps better deferred to Part II, where its relationship to de Rham cohomology will be made clear, nonetheless, some preliminary remarks about the nature of work as a linear functional on 1-chains that take the form of straight line segments can be made. One must simply restrict oneself to applied forces that are constant along the displacement to begin with.

When dealing with forces that act along finite displacements, one must realize that one is automatically dealing with displacement as a motion in space, and not merely a geometric relationship between vertices. In particular, the internal force in a rigid link will produce no displacement, and thus do no work. Hence, it is typically only the external forces that might possibly do work in a rigid mechanical network, and the only finite displacement that would be consistent with that rigidity would be a rigid motion. However, when one allows deformable links or when the network has rigid links but internal degrees of freedom, one must also consider non-rigid motions of the nodes and links. In the terminology that was introduced before in this treatise, only the kinematical complex that is defined by the motion (viz., a vector homology) will be associated with work, not the successive structural complexes.

Suppose that a constant force $F$ acts along a straight line segment $\{x\,(t), t_0 \leq t \leq t_1\}$ in an affine space $A^n$, which is then the geometric realization of a 1-simplex $\bar{\sigma}_1$. If that line segment $\bar{\sigma}_1$ is associated with the displacement vector in $\mathbb{R}^n$:

$$\mathbf{u} = \Delta\mathbf{x} = x\,(t_1) - x\,(t_0) \qquad (8.9)$$

then the work that is done by $F$ during that displacement will be the real number:

$$<W\,(F),\ \bar{\sigma}_1> = <F, \Delta\mathbf{x}> = <F, \mathbf{x}\,(t_1) - \mathbf{x}\,(t_0)>. \qquad (8.10)$$

If we promote $F$ to the 1-simplex $F\,\bar{\sigma}_1$ with coefficients in $\mathbb{R}^{n^*}$ and represent the displacement vector as the reciprocal 1-cosimplex $\Delta\mathbf{x}\ \bar{\sigma}^1$ with values in $\mathbb{R}^n$ then:

$$<W\,(F),\ \bar{\sigma}_1> = <\Delta\mathbf{x}\ \bar{\sigma}^1,\ F\,\bar{\sigma}_1> = <F, \Delta\mathbf{x}><\bar{\sigma}^1,\ \bar{\sigma}_1> = <F, \Delta\mathbf{x}>. \qquad (8.11)$$

From (8.11), one can think of the work done by $F$ in this case as a 1-cochain with values in $\mathbb{R}$, namely:

$$W(F) = \; <F, \Delta \mathbf{x}> \bar{\sigma}^1 .\qquad (8.12)$$

If our elementary kinematical complex $\mathcal{K}$ is generated by $\{ \sigma_0(0), \sigma_0(1), \bar{\sigma}_1 \}$ then we can say that $W(F) \in C^1(\mathcal{K}; \mathbb{R})$.

In order for $W(F)$ to vanish, in addition to the vanishing of $F$ or $\Delta \mathbf{x}$, another sufficient condition would be that the vector $\mathbf{F}$ that is associated with the covector $F$ by way of the Euclidian scalar product is perpendicular to the displacement vector $\Delta \mathbf{x}$. That is why forces of constraint will do no work when the constraints are perfect (i.e., normal to the constraint manifold, which would not be true if there were friction).

Since $F$ is constant along $\bar{\sigma}_1$, we can also say that:

$$W(F) = [<F, \mathbf{x}(t_1)> - <F, \mathbf{x}(t_0)>] \; \bar{\sigma}^1 = -\Delta U, \qquad (8.13)$$

when we define:

$$U = U(0) \; \sigma^0(0) + U(1) \; \sigma^0(1), \qquad U(A) \equiv -<F, \mathbf{x}(t_A)>, \qquad A = 0, 1. \qquad (8.14)$$

We can check this by evaluating $<\Delta U, \bar{\sigma}_1> = \; <U, \partial \bar{\sigma}_1> = \; <U, [\sigma_0(1) - \sigma_0(0)]>$ . Hence, we can regard the 0-cochain $U$ as a "potential function" for $W(F)$.

Now, let us look at a more general case, in which there are $r_0$ initial nodes $\sigma_0(i)(0)$, $i = 1, \ldots, r_0$, and:

$$\bar{F} = \sum_{i=1}^{r_0} F(i) \bar{\sigma}_1(i), \qquad \Delta \mathbf{x} = \sum_{i=1}^{r_0} \Delta \mathbf{x}(i) \bar{\sigma}^1(i) . \qquad (8.15)$$

The overbar on $F$ is intended to distinguish the 1-chain that the $F(i)$ define from the corresponding 0-chain.

Furthermore, the kinematical complex $\mathcal{K}$ now represents an elementary vector homology between two 1-complexes, so in particular, the 1-simplexes $\bar{\sigma}_1(i)$ will all connect the 0-simplexes $\sigma_0(i)(0)$ in the initial complex to 0-simplexes $\sigma_0(i)(1)$ in the final complex; i.e., $\partial \bar{\sigma}_1(i) = \sigma_0(i)(1) - \sigma_0(i)(0)$ . Those 1-simplexes, together with the initial and final 0-simplexes, and the boundary operator that links them constitute the kinematical complex that the forces and displacements are associated with.

The natural bilinear pairing $\mathbb{R}^{n^*} \times \mathbb{R}^n \to \mathbb{R}$, $(\alpha, \mathbf{v}) \mapsto \alpha(\mathbf{v})$ can be used to define bilinear pairings $C^0(\mathcal{K}; \mathbb{R}^{n^*}) \times C_0(\mathcal{K}; \mathbb{R}^n) \to \mathbb{R}$ and $C^1(\mathcal{K}; \mathbb{R}^{n^*}) \times C_1(\mathcal{K}; \mathbb{R}^n) \to \mathbb{R}$ . In particular, the total work done by $\bar{F}$ along all of the displacements is:

$$<\Delta \mathbf{x}, \bar{F}> = \sum_{i=1}^{r_0} \sum_{j=1}^{r_0} <\Delta \mathbf{x}(i), F(j)> <\bar{\sigma}^1(i), \bar{\sigma}_1(j)> = \sum_{i=1}^{r_0} <F(i), \Delta \mathbf{x}(i)> . \qquad (8.16)$$

Since the 1-chain $\overline{c}_1 = \sum\limits_{i=1}^{r_0} \overline{\sigma}_1(i)$ is the carrier of the displacement 1-cochain $\Delta\mathbf{x}$, we will define the work done by $\overline{F}$ along $c_1$ to be:

$$< W(F),\, \overline{c}_1 > = <\Delta\mathbf{x},\, \overline{F} > . \tag{8.17}$$

Hence, by rearranging the middle equation in (8.16), we can express the real-valued 1-cochain $W(F)$ as:

$$W(F) = \sum\limits_{i=1}^{r_0} < F(i), \Delta\mathbf{x}(i) > \overline{\sigma}^1(i)\,. \tag{8.18}$$

In order for this to vanish, $< F(i), \Delta\mathbf{x}(i) >$ must vanish for every $i$, which reverts to the aforementioned sufficient conditions for $F(i)$ and $\Delta\mathbf{x}(i)$ .

Since there are no 2-cochains in our complex except for 0, we must have that $\Delta W(F) = 0$, so $W(F)$ must be a 1-cocycle.  Hence, the work done by $F$ around any 1-boundary must vanish, but the only 1-boundary is 0.

The question then arises whether $W(F)$ is also a 1-coboundary, which would make the work done around any 1-cycle $\overline{z}_1$ vanish.  However, at this point, our kinematical complex does not have any 1-cycles, except for 0.  Nonetheless, by definition, $W(F)$ will be a 1-coboundary iff there is some 0-cochain ([1]):

$$U = \sum\limits_{A=0}^{1} \sum\limits_{i=1}^{r_0} U(i)(A)\, \sigma^0(i)(A) \tag{8.19}$$

such that:

$$W(F) = -\Delta U(F) = -\sum\limits_{i=1}^{r_0} \Delta U(i)\, \overline{\sigma}^1(i) = -\sum\limits_{i=1}^{r_0} \big[U(i)(1) - U(i)(0)\big]\overline{\sigma}^1(i)\,; \tag{8.20}$$

i.e.:

$$W(F)(i) = < F(i), \Delta\mathbf{x}(i) > = -\,[U(i)(1) - U(i)(0)]\ \text{for all } i. \tag{8.21}$$

By analogy with (8.14), we find that we can define:

$$U(i)(A) = -< F(i), \mathbf{x}(i)(A) > . \tag{8.22}$$

We shall then call the 0-cochain $U(F)$ a *potential energy function* for $W(F)$.  The reason that we say "a potential energy function," instead of "*the* …" is because it is not unique.  Any two $U(F)$ and $U'(F)$ that differ by a 0-cocycle will yield the same $W(F)$.  Such a 0-cocycle will take the form:

$$\overline{z}^0 = \sum\limits_{A=0}^{1} \sum\limits_{i=1}^{r_0} a(i)\, \sigma^0(i)(A)\,, \tag{8.23}$$

---

[1]   From now on, we shall abbreviate the time point $t_A$ to its index $A$.

in which the $a(i)$ are real constants.

Whether or not $W(F)$ admits such a $U(F)$ obviously depends upon the connectivity of the complex $\mathcal{K}$; i.e., $H^1(\mathcal{K};\mathbb{R})$. The existence of such a $U(F)$ is equivalent to demanding that $<W(F),\ \overline{z}_1>$ will vanish whenever $\overline{z}_1$ is a 1-cycle, so when $H^1(\mathcal{K};\mathbb{R})$ is not zero, there will be 1-cycles $\overline{z}_1$ for which $<W(F),\ \overline{z}_1>$ is non-vanishing.

However, the existence of a $U(F)$ also depends upon the nature of $F$. Since we have restricted ourselves to forces that are constant along the displacement, we have obtained a potential function automatically. However, in the case of the force of friction, the force vector is always proportional to the displacement vector, which can change in direction. Moreover, the work done will be proportional to the length of each displacement, which will be positive in any non-trivial case. As long as the total length of the cycle is non-zero, the work done will also be non-zero.

More generally, if we go back to (8.16), we will also see that:

$$<\Delta\mathbf{x},\overline{F}> = <\mathbf{x},\partial\overline{F}>,\tag{8.24}$$

which will vanish for all $\mathbf{x}$ iff $\partial\overline{F}=0$; i.e., $\overline{F}$ is a 1-cycle.

Of course, with our original definition (8.15) of $\overline{F}$, we will have only:

$$\partial\overline{F}=\sum_{i=1}^{r_0}F(i)\,\partial\overline{\sigma}_1(i)\ =\ \sum_{i=1}^{r_0}F(i)[\sigma_0(i)(1)-\sigma_0(i)(0)],\tag{8.25}$$

which will vanish iff for all $i$, either $F(i)=0$ or $\sigma_0(i)(0)=\sigma_0(i)(1)$; i.e., either the force on the node $i$ vanishes or it does not displace.

In order to have non-trivial 1-cycles that are associated with the action of forces, we must basically concatenate vector homologies and then look at the spatial trace of the space-time complex, as we discussed before. Mostly, that extension of our kinematical complex involves allowing the number of time points to extend from $A=0$ to $A=N$, which is some positive integer that is typically greater than 1. Hence, our new force 1-chain and displacement 1-cochain will now take the forms:

$$\overline{F}\ =\ \sum_{A=0}^{N}\sum_{i=1}^{r_0}F(i)(A)\,\overline{\sigma}_1(i)(A),\qquad\Delta\mathbf{x}=\sum_{A=0}^{N}\sum_{i=1}^{r_0}\Delta\mathbf{x}(i)(A)\,\overline{\sigma}^1(i)(A),\tag{8.26}$$

respectively.

The total work done by $\overline{F}$ over all displacements in $\Delta\mathbf{x}$ will then be:

$$<\Delta\mathbf{x},\ \overline{F}>=\sum_{A=0}^{N}\sum_{i=1}^{r_0}\sum_{j=1}^{r_0}<\Delta\mathbf{x}(i)(A),F(j)(A)><\overline{\sigma}^1(i)(A),\tilde{\sigma}_1(j)(A)>\tag{8.27}$$

$$=\sum_{A=0}^{N}\sum_{i=1}^{r_0}<F(i)(A),\Delta\mathbf{x}(i)(A)>.\tag{8.28}$$

By rearranging the summations, we can define the work done by $F$ to be the 1-cochain:

$$W(F) = \sum_{A=0}^{N} \sum_{i=1}^{r_0} < F(i)(A), \Delta \mathbf{x}(i)(A) > \bar{\sigma}^1(i)(A). \tag{8.29}$$

The vanishing of $W(F)$ is now more involved than simply the vanishing of each coefficient of $\bar{\sigma}^1(i)(A)$ individually, since we now have the possibility of loops around which the work might also vanish without the individual contributions from each 1-simplex in the path being zero. Hence, let us first reorganize the sum in $W(F)$ to reflect the fact that each initial vertex $i$ is associated with associated with a path and a total displacement:

$$l(i) = \sum_{A=0}^{N} \bar{\sigma}_1(i)(A), \qquad \Delta \mathbf{x}(i) = \sum_{A=0}^{N} \Delta \mathbf{x}(i)(A) \bar{\sigma}^1(i)(A) \tag{8.30}$$

resp., so the total work done by $F$ along the path $l(i)$ will be:

$$< W(F), l(i) > = <\Delta \mathbf{x}(i), \bar{F} > = \sum_{A=0}^{N} < F(i)(A), \Delta \mathbf{x}(i)(A) >, \tag{8.31}$$

and the total work done by $F$ over all displacements will be simply:

$$W(F) = \sum_{i=1}^{r_0} < W(F), l(i) >. \tag{8.32}$$

We can now concentrate on the work done along each path $i$.

For each path $i$, the corresponding 1-cochain take the form:

$$W(F)(i) = \sum_{A=0}^{N} < F(i)(A), \Delta \mathbf{x}(i)(A) > \bar{\sigma}^1(i)(A). \tag{8.33}$$

It is now entirely possible for $W(F)(i)$ to vanish without all of the coefficients of each $\bar{\sigma}^1(i)(A)$ vanishing, as well. In order to prove that, we first show that $W(F)(i)$ is a 1-coboundary.

Once again, $W(F)(i)$ will still be a 1-cocycle, and for the same reason. As long as the individual forces $F(i)(A)$ are constant along the displacements, it will also be a 1-coboundary when we define the potential energy 0-cochain to be:

$$U(i) = \sum_{A=0}^{N} U(i)(A) \sigma^0(i)(A), \tag{8.34}$$

with

$$U(i)(A) = - <F(i)(A), \Delta \mathbf{x}(i)(A) >. \tag{8.35}$$

Once again, which is not unique, but is defined only up to a 0-cocycle, which will then take the form:

$$\overline{z}^0(i) = \sum_{A=0}^{N} a(i) \sigma^0(i)(A).$$  (8.36)

Hence, from the basic property of 1-coboundaries, we can assert:

**Theorem:**

*The work done by F around any 1-cycle will vanish.*

We can now define a force distribution *F* to be *conservative* iff the work done by it around any 1-cycle vanishes. That immediately suggests the following:

**Theorem:**

*The following are equivalent:*

1. *F is conservative.*

2. $W(F)(\overline{z}_1) = 0$ *for any 1-cycle* $\overline{z}_1$ .

3. $W(F)$ *is a 1-coboundary.*

4. $W(F)$ *is independent of the path between two vertices.*

**Proof:**

The first three equivalences follow from definitions.

If $l_1$ and $l_2$ are two paths that connect two successive vertices $\sigma_0(0)$ and $\sigma_0(1)$ in time, so $\partial l_1 = \partial l_2 = \sigma_0(1) - \sigma_0(0)$ then $l_2 - l_1$ will be a 1-cycle. Hence, $W(F)(l_2 - l_1)$ will vanish, but:

$$W(F)(l_2 - l_1) = W(F)(l_2) - W(F)(l_1),$$

from the linearity of $W(F)$. Since that must vanish, we will have:

$$W(F)(l_2) = W(F)(l_1)$$

for any such $l_1$ and $l_2$ .

*c. Power.* – When a resultant force *F* (*i*) acts upon a node $\sigma_0$ (*i*) of a structural complex that moves with a velocity **v** (*i*), the power that is added or dissipated along the associated link $\overline{\sigma}_1(i)$ of the kinematical complex will be the real number:

$$P(i) = <F(i), \mathbf{v}(i)> .$$  (8.37)

If we extend this to a time sequence $t_A$, $A = 0, \ldots, N$:

$$P\,(i)(A) = <F\,(i)(A)\,,\,\mathbf{v}\,(i)(A)> \tag{8.38}$$

then we can assemble the contributions from all such links into a real-valued 1-cochain:

$$P = \sum_{A=0}^{N} \sum_{i=1}^{r_0} P(i)(A)\,\bar{\sigma}^1(i)(A)\,. \tag{8.39}$$

that represents the total power that is added or dissipated by all links in the kinematical complex.

If each force $F\,(i)(A)$ is constant in time over the link $\bar{\sigma}_1(i)(A)$ then since $\mathbf{v}\,(i)(A) = d\,(\mathbf{x}\,(i)(A))\,/\,dt$, one will have:

$$P\,(i)(A) = \frac{d}{dt} <F\,(i)(A),\,\mathbf{x}\,(i)(A)>\,. \tag{8.40}$$

When one integrates $P\,(i)(A)$ over time along the 1-simplex $\bar{\sigma}_1(i)$, one will get the real number:

$$\int_{t_0}^{t_1} P(i)(A)\,dt = \big[<F(i)(A),\mathbf{x}(i)(A)>\big]_{t_0}^{t_1} = <F\,(i)(A),\,\Delta\mathbf{x}\,(i)(A)> \text{‘}$$

i.e.:

$$\int_{t_0}^{t_1} P(i)(A)\,dt = W\,(F)\,(i)(A)\,. \tag{8.41}$$

which implies that:

$$P\,(i)(A) = \frac{d}{dt}\,W\,(F)\,(i)(A)\,. \tag{8.42}$$

Hence, power is the time rate of doing work, in this restrictive case of forces that are constant along the links of the kinematical complex. (We will generalize in Part II.)

*d. The work-kinetic energy theorem.* – If one also forms $<dp\,(i)\,/\,dt,\,\mathbf{v}\,(i)>$ for a vertex $\sigma_0\,(i)$ in a structural complex then in the event that the linear momentum $p\,(i)$ is convective [so it will take the form $m\,(i)\,v\,(i)$] and $m\,(i)$ is constant in time, one will have:

$$<\frac{dp(i)}{dt},\mathbf{v}(i)> = \frac{d}{dt}\Big[\tfrac{1}{2}m(i)\,v(i)^2\Big] = \frac{d}{dt}\,\text{KE}\,(i)\,. \tag{8.43}$$

When one integrates this over time along a straight line segment that carries the 1-simplex $\bar{\sigma}_1(i)$, one will get the real number:

$$\Delta\text{KE}\,(i) = \int_{t_0}^{t_1}\Big[\frac{d}{dt}\text{KE}(i)\Big]dt = \text{KE}\,(i)(1) - \text{KE}\,(i)(0)\,. \tag{8.44}$$

If one extends the single time interval $[t_0, t_1]$ to a time sequence $t_A$, $A = 0, \ldots, N$ then one can assemble the contributions from the links $\Delta KE$ $(i)(A)$ into the real-valued 1-cochain:

$$\Delta KE = \sum_{A=0}^{N} \sum_{i=1}^{r_0} \Delta KE\,(i)(A)\,\sigma^1(i)(A)\,, \qquad (8.45)$$

which will represent the change in the total kinetic energy over the motion of the initial kinematical complex. In fact, if we choose the path $l$ $(i)$ that is followed by a chosen $\sigma_0$ $(i)$, in time then:

$$\Delta KE\,(i) = \sum_{A=0}^{N} \Delta KE\,(i)(A)\,\bar{\sigma}^1(i)(A) \;=\; \Delta\left[\sum_{A=0}^{N} KE\,(i)(A)\,\sigma^0(i)(A)\right]$$

$$= [KE\,(i)(N) - KE\,(i)(0)]\ \bar{\sigma}_1(i)(0)\,.$$

In other words:

**Theorem:**

*The total change in kinetic energy along any path will be the difference between its final and initial values.*

When one sums these path contributions over all paths, one will get an analogous statement for the total change in kinetic energy:

$$\Delta KE \equiv \sum_{i=1}^{r_0} \Delta KE\,(i)\,. \qquad (8.46)$$

When Newton's second law is applied to the expression $< dp\,/\,dt,\ \mathbf{v} >$ that will imply the:

**Work-kinetic energy theorem:**

*When a force distribution is conservative and the linear momentum is convective with masses that are constant in time over each line of the kinematical complex of a motion, one will have:*

$$W\,(F) = \Delta\,KE\,. \qquad (8.47)$$

Since $W\,(F) = -\,\Delta U\,(F)$ for a conservative force distribution, this will become:

$$\Delta\,KE + \Delta\,U\,(F) = 0, \qquad (8.48)$$

and if we define the *total energy* by:

$$E_{\text{tot}} = KE + U\,(F), \qquad (8.49)$$

which will then be a real-valued 0-cochain, then we will also have:

$$\Delta E_{\text{tot}} = 0, \tag{8.50}$$

which will also make $E_{\text{tot}}$ a 0-cocycle.

Since that also represents the change in the total energy over the time interval of the motion, we can assert the following extension of a previous theorem:

**Conservation of energy theorem:**

*The following are equivalent:*

i)  *The force distribution is conservative.*

ii)  *The work functional $W(F)$ admits a potential energy $-U(F)$.*

iii)  *$W(F)$ is a 1-coboundary.*

iv)  $\Delta \text{KE} = -\Delta U(F)$.

v)  *$E_{\text{tot}}$ is a 0-cocycle.*

vi)  *$E_{\text{tot}}$ is constant in time.*

**9. The principle of virtual work and d'Alembert's theorem ([1]).** – Although defining the concept of a finite amount of work required an actual finite motion to take place, one can still introduce a concept of work in statics that has an infinitesimal character, namely, virtual work, which is the work done by a force distribution under an infinitesimal deformation of the mechanical network that is called a "virtual displacement." Not surprisingly, much of what one considers runs parallel to a discussion of the calculus of variations, but that is because a virtual displacement is essentially the same thing as a variation. However, the virtual work is not actually analogous to the action functional, but to its first variation. As a result virtual work can be defined in cases where action functionals do not exist, such as non-conservative forces and imperfect or non-holonomic constraints. (See the author's discussion of this in [**22**].)

   *a. Virtual displacements of a mechanical network.* – A *virtual displacement* of the nodes and branches of a structural complex differs from a real displacement by the fact that it is really an infinitesimal generator of a one-parameter family of real displacements, and can then be regarded as a set of vectors in $\mathbb{R}^n$ that are associated with each node and link. One can also think of it as a "variation" $\delta \mathbf{x}$ of the state that is defined by the position distribution $\mathbf{x}$. Furthermore, the resulting change $\delta \mathbf{s}$ in the displacement

---

[1]  A good reference on the physics of the topic of this section is Sommerfeld [**21**].

distribution **s** must be coupled to the change in the position distribution by the relationship:

$$\delta \mathbf{s} = \Delta (\delta \mathbf{x}). \tag{8.51}$$

(This can be proved by differentiating the one-parameter families.) Hence, $\delta \mathbf{x} \in C^0(\mathcal{L}_1 ; \mathbb{R}^n)$ and $\delta \mathbf{s} \in C^1(\mathcal{L}_1 ; \mathbb{R}^n)$.

Recall that our earlier definition of an infinitesimal motion of the network was a velocity vector field that was defined on the nodes, which we pointed out was the generator of a one-parameter family of finite motions. Hence, one can also think of a virtual displacement as an infinitesimal motion, although more precisely, one has: $\delta \mathbf{x} = \delta t$.

*b. The principle of virtual work.* – When one evaluates the external force 0-cochain $F$ that acts upon a structural complex on a virtual displacement $\delta \mathbf{s}$, one will get:

$$< F, \delta \mathbf{s} > = \sum_{a=1}^{r_1} \sum_{b=1}^{r_1} < F(a), \delta \mathbf{s}(b) > < \sigma^1(a), \sigma_1(b) > = \sum_{a=1}^{r_1} < F(a), \delta \mathbf{s}(a) >, \tag{8.52}$$

which will then represent the *total virtual work* that is done by the virtual displacement of the configuration as a result of the force, and which will be denoted by $\delta W (F)(\delta \mathbf{s})$.

One immediately sees from (8.51) that:

$$\delta W (F)(\delta \mathbf{s}) = < F, \Delta (\delta \mathbf{x}) > = < \partial F, \delta \mathbf{x} >. \tag{8.53}$$

That will vanish for all $\delta \mathbf{x}$ iff $\partial F = 0$. However, that is also the condition for force equilibrium.

**Theorem (principle of virtual work):**

*A mechanical network is in force equilibrium iff the virtual work that is done by F on any virtual displacement is zero.*

*c. D'Alembert's principle.* – Although the principle of virtual work seems to be limited to problems of statics, it was the insight of d'Alembert that one could still use it as a principle of dynamics. All that was required was to associate the motion itself with an "force of inertia." As a result, one could regard the natural state of motion for a system as being essentially an equilibrium state in space-time, rather than merely in space.

An easy example of how one might conceptualize that force of inertia is given by a cable lifting a weight. The total tension $T$ in the cable is going to come from both the weight itself $W = mg$ and the mass $m$ that is associated with $W$ times the (signed) acceleration $a$ of the weight during that motion, so $T = W \pm ma = m (g \pm a)$. The positive sign refers to motion upward, while the negative sign refers to downward motion. Hence, in a sense, one can think of the force that *ma* represents as the force of inertia in this case. More generally, if $p$ is the linear momentum of the mass then one could also use $dp / dt$.

Hence, if $\overline{F}$ is the force 1-chain that is associated with a virtual displacement $\Delta \delta \mathbf{x}$ and the displacement is associated with a momentum 0-chain $p$ whose time derivative is $dp / dt$ and which extends to a 1-chain $\overline{dp} / dt$ then:

**Theorem (d'Alembert's principle):**

*The natural motion of a mechanical network that is acted upon by external forces is the one for which:*

$$\delta W = \left< \left( \overline{F} - \frac{dp}{dt} \right), \Delta(\delta \mathbf{x}) \right> = 0 \qquad (8.54)$$

*for every virtual displacement $\Delta \delta \mathbf{x}$.*

In order to see that, consider that:

$$\left< \left( \overline{F} - \frac{dp}{dt} \right), \Delta(\delta \mathbf{x}) \right> = \left< \partial \overline{F} - \partial \frac{dp}{dt}, \delta \mathbf{x} \right>,$$

which will vanish for all $\delta \mathbf{x}$ iff:

$$\partial \overline{F} = \partial \frac{dp}{dt} = \frac{d}{dt} \partial \overline{p},$$

which is basically Newton's second law, since $\partial \overline{F}$ will give the resultant force that acts upon each node of the network, while $\partial \overline{p}$ will give the linear momentum of that node.

**10. Open and closed networks.** One of the recurring themes of physical systems amounts to their completeness, in the sense of whether all of the factors that determine their state are accounted for the in the mathematical model. If so, one thinks of the systems as being "closed," and if not, "open."

Generally, the issue is one of where one draws the imaginary "box" that separates the "internal" structure of the system from the "external" structure, which one usually imagines to be simplified in some way. For instance, external influences are assumed to be beyond the influence of the state of the system, such as when a heat bath or reservoir for an energetic system does not change in temperature as a result of its energetic interaction with the system.

It is in the examples where one deals with the interactions of two or more simple systems that have a comparable sort of magnitude (in some sense) that one finds that the distinction between internal and external might be debatable. For instance, the orbiting of the Earth around the Sun can be modeled to a reasonable degree of approximation by assuming that the mass of the Sun is infinite, so that its own motion in response to the mutual force of gravitation between the two masses can be ignored, but when one is considering a double star system for which both stars have comparable masses, that approximation would no longer be justifiable.

Similarly, when one steps off of a small boat and has to contend with the way that the boat moves in the opposite direction to one's own motion, one can draw an imaginary box around the boat and say that the force of interaction between one's foot and the boat is external to the boat, so it will accelerate and its momentum will increase during that interaction. Of course, a similar statement can be made for your own motion. However, if one draws the box around both oneself and the boat then the interaction force will be internal to the total system, and as a result the total momentum will be conserved throughout the process.

For a system of interacting masses that is described by a structural complex, the distinction between internal and external basically comes down to a partitioning of the sets of vertices $V$ and branches $B$ of the network into disjoint unions of sets $V_i$ and $V_e$, $B_i$ and $B_e$, resp., that represent the internal and external branches and vertices, respectively. Although the partitioning of $V$ can be arbitrary, for the sake of consistency, one must say that a branch $b$ is *internal* iff $\partial b$ consists of two internal vertices, while it is *external* otherwise. One can further distinguish between *completely external* branches, for which both incident vertices are external, and *linking* branches, for which one is internal and the other is external. In many cases, the external vertex of a linking branch will not be incident on any other branches, and we will call such a link *isolated*.

As we said before, if the internal nodes of the network are all in equilibrium, then we will find that:

$$\partial F = \sum_{i=1}^{r_{0,e}} F(i)\, \sigma_0^e(i) = -F_{\text{ext}}, \tag{9.1}$$

in which $r_{0,e}$ represents the number of external nodes.

In the statics of discrete structures, one must insure that the resultant of the external forces is zero, in addition to the resultants at each of the internal nodes. Moreover, the external nodes are often all isolated. Here, one finds that it is convenient to think of all the isolated external branches as being incident on the same "point at infinity" $P_\infty$, which we also represent by the 0-simplex $\sigma_0(\infty)$. That is, we are imposing the equivalence relation upon the vertices or 0-simplexes that:

$$\sigma_0(\infty) = \sigma_0^e(1) = \ldots = \sigma_0^e(r_{0,e}) \ .$$

As a result of this equivalence, the second equality in (9.1) will become:

$$F_{\text{ext}} = \left( \sum_{i=1}^{r_{0,e}} F(i) \right) \sigma_0(\infty) \,, \tag{9.2}$$

and the total system will be in equilibrium iff the resultant of all external forces vanishes, as well as the internal ones.

Of course, the fact that $P_\infty$ is at infinity also implies that the displacements $P_\infty - \mathbf{x}_i$ that are associated with the isolated external branches are effectively infinite in length. In some cases, that will not be fatal, such as when the force of interaction is inversely proportional to some positive power of that distance and thus vanishes at infinity.

The device of identifying all isolated external vertices with a single point at infinity amounts to a "one-point compactification" of the network. This is not precisely the same as when one compactifies the space in which the network lives, such as when one turns a planar graph into a spherical graph by compactifying the plane with a point at infinity that generally does not belong to the vertices of the graph, through.

## References (*)

---

(*) References that are marked with an asterisk are available in English translation as free PDF downloads from the author's website.

      

___________